\documentclass{article}

\usepackage[square,sort&compress,comma,numbers]{natbib}
\usepackage[preprint]{neurips_2021}

\usepackage[utf8]{inputenc} 
\usepackage[T1]{fontenc}    
\usepackage{hyperref}       
\usepackage{url}            
\usepackage{booktabs}       
\usepackage{amsfonts}       
\usepackage{nicefrac}       
\usepackage{xcolor}         

\usepackage{changes}
\definecolor{blue}{rgb}{1.,0,0}

\usepackage{float}

\usepackage{tikz}
\usepackage{amsmath, amssymb}
\usepackage{kotex}
\usepackage{graphicx}
\graphicspath{{images/}}
\usepackage{siunitx}
\usepackage{xspace}
\usepackage{lipsum}
\usepackage{subcaption}
\usepackage{multirow}
\usepackage{pbox}
\usepackage{tabularx}
\usepackage{verbatim}
\usepackage{pifont}
\usepackage{caption} 

\title{Controllable and Interpretable \\ Singing Voice Decomposition via Assem-VC}

\author{%
  Kang-wook Kim \\
  MINDsLab Inc., Seoul National University \\
  \texttt{full324@snu.ac.kr} \\
  \And
  Junhyeok Lee \\
  MINDsLab Inc. \\
  \texttt{jun3518@mindslab.ai} \\
}

\makeatletter
\DeclareRobustCommand\onedot{\futurelet\@let@token\@onedot}
\def\@onedot{\ifx\@let@token.\else.\null\fi\xspace}

\def\eg{\emph{e.g}\onedot}

\def\etal{\emph{et al}\onedot}
\makeatother

\newcommand{\mel}{mel spectrogram\xspace}

\begin{document}

\maketitle

\section{Introduction}

Singing is the most accessible among musical arts, to people of all abilities.
Following the success of its applications in various fields, deep neural networks have also been applied to singing voice synthesis.
While prior works \cite{lee2019adversarially, choi2020korean, mlpsinger, nsinger} attempted to synthesize singing voices from linguistic, melodic, and temporal information, they heavily relied on time-aligned phonemes and pitches, forcing users to control down to the smallest details. 
Thus, people who can utilize such tools are limited to those who are familiar with music scores and musical instrument digital interface (MIDI).
This bars the general public without musical expertise from express their creativity.

Recently, there have been efforts \cite{li2020ppg, ppg_svs, liu2020fastsvc, polyak2020unsupervised} to convert singing voices to other singer's voices.
However, since most singing voice conversion models use phonetic posteriorgrams (PPG) \cite{li2020ppg, ppg_svs} or map the linguistic content into uninterpretable latent codes \cite{liu2020fastsvc, polyak2020unsupervised}, they cannot control any phoneme-level information.

Allowing the users to easily control lyrics, rhythm, pitch, and timbre of a singing voice would open the door to a new musical application. 
For example, the users can modify an existing singing voice to their needs, or convert the existing singing voice to a different voice.
Moreover, with timbre conversion, the user can control the target singer's voice without explicit time-aligned input such as MIDI.
In this paper, we propose a singing voice decomposition system that encodes the four attributes of a singing voice: linguistic content, rhythm, pitch, and speaker identity.
To the best of our knowledge, this is the first work to disentangle and control each attribute of a singing voice.

\section{Singing Voice Decomposition System}
\begin{figure}[t]
    \centering
    \includegraphics[width=\linewidth]{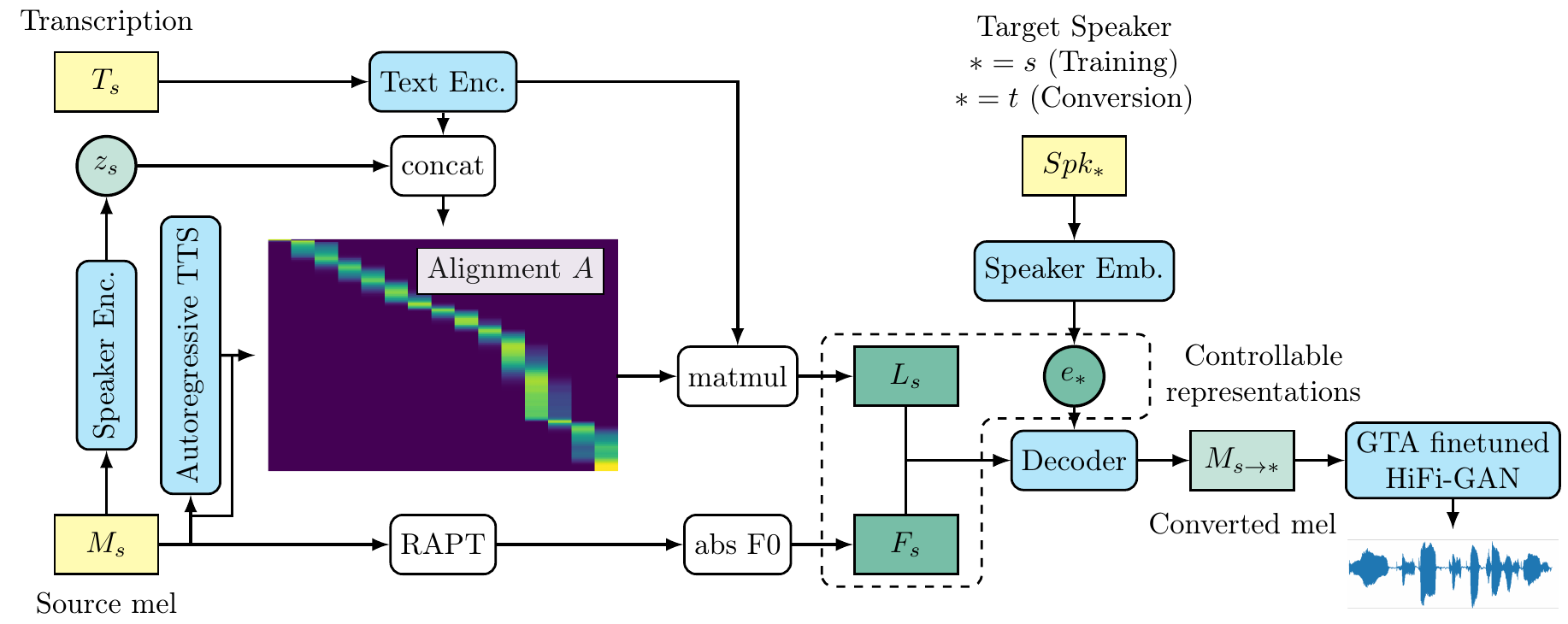}
    \caption{Overall architecture and attributes. Yellow blocks indicate the input of the system, green blocks denote the encoded attributes and blue blocks denote learnable modules.}
    \label{fig:overall}
\end{figure}

Our goal is to encode linguistic content, rhythm, pitch, and timbre as interpretable formats, which enables easy control of singing voices without the need for temporal-aligned and detailed inputs \eg{} duration labels for phonemes or pitches or MIDI.
We apply the state-of-the-art many-to-many voice conversion system, Assem-VC\footnote{Pre-trained weights and code of Assem-VC are available on  \href{https://github.com/mindslab-ai/assem-vc}{https://github.com/mindslab-ai/assem-vc}} \cite{assem-vc} to singing voice decomposition.
Our system takes the source singing voice and the corresponding lyrics as the input and estimates the alignment between the \mel and the transcript of the singing voice.
Lyrics are converted to phoneme sequences with a proprietary grapheme-to-phoneme (G2P) algorithm.
We use Cotatron \cite{cotatron} and RAPT \cite{rapt} as the alignment encoder and the fundamental frequency (F0) estimator, respectively, of the singing voice.
Since this approach deals with singing voices rather than ordinary speech, the absolute sequence of F0 is used over the logarithm of the speaker-normalized pitch for additional interpretability.
We use speaker embedding rather than speaker encoder for speaker conditioning in the decoder since empirically, the speaker encoder tends to overfit in data-hungry situations.
Fig. \ref{fig:overall} shows the overall architecture and the attributes encoded with our architecture.
Details of training the entire system including the HiFi-GAN vocoder \cite{hifigan} are described in Appendix \ref{details}.

\begin{figure*}[th!]
\setlength{\textfloatsep}{0pt}%

\newcommand{\incaptionimg}[3]{
  \begin{tikzpicture}[every node/.style={inner sep=0,outer sep=0}]
    \draw node[name=micrograph] {\includegraphics[width=\textwidth]{#1}}; 
    \draw  (micrograph.north west)  node[anchor=north west,yshift=-0.1cm,xshift=0.1cm,#3]{\textbf{\small{(#2)}}}; 
  \end{tikzpicture}
}
\newcommand{\incaptionlineimg}[4]{
  \begin{tikzpicture}[every node/.style={inner sep=0,outer sep=0}]
    \draw node[name=micrograph] {\includegraphics[width=\textwidth]{#1}}; 
    \draw  (micrograph.north west)  node[anchor=north west,yshift=-1.07cm,xshift=0.44cm,#3]{\textbf{\small{(#2)}}}; 
  \end{tikzpicture}
}
\captionsetup[subfigure]{labelformat=empty}
    \setlength{\textfloatsep}{0pt}%
    \setlength{\intextsep}{0pt}
    \centering
    \begin{subfigure}[b]{0.25\linewidth}
    \subcaption{Reference Signal}
    \end{subfigure}%
    \begin{subfigure}[b]{0.25\linewidth}
    \subcaption{Linguistic Content}
    \end{subfigure}%
    \begin{subfigure}[b]{0.25\linewidth}
    \subcaption{Rhythm}
    \end{subfigure}%
    \begin{subfigure}[b]{0.25\linewidth}
    \subcaption{Pitch}
    \end{subfigure}\\
    \vspace{-0.80\baselineskip}
    \begin{subfigure}[b]{0.25\linewidth}
         \incaptionimg{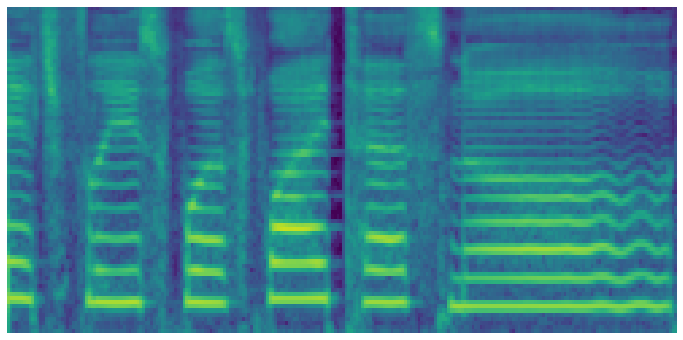}{a}{white}
         \phantomsubcaption\ignorespaces\label{spec_f}
     \end{subfigure}%
     \begin{subfigure}[b]{0.25\linewidth}
         \incaptionimg{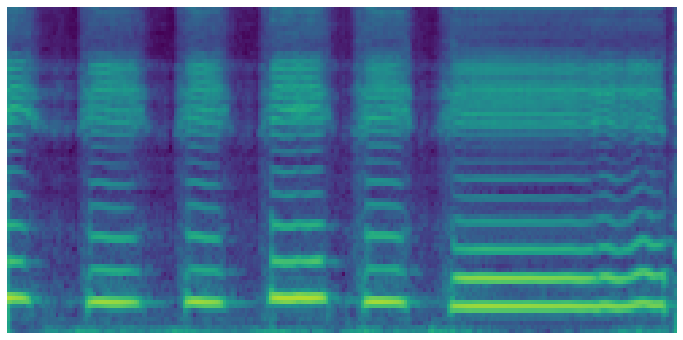}{b}{white}
         \phantomsubcaption\ignorespaces\label{spec_g}
     \end{subfigure}%
     \begin{subfigure}[b]{0.25\linewidth}
         \incaptionimg{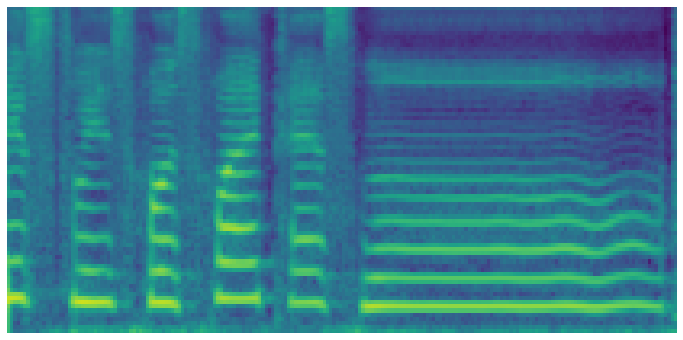}{c}{white}
         \phantomsubcaption\ignorespaces\label{spec_h}
     \end{subfigure}%
     \begin{subfigure}[b]{0.25\linewidth}
         \incaptionimg{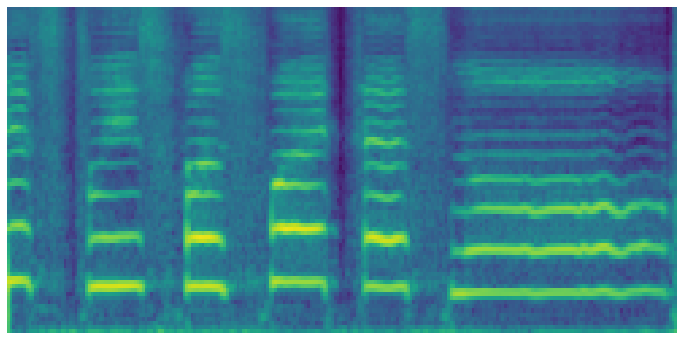}{d}{white}
         \phantomsubcaption\ignorespaces\label{spec_j}
     \end{subfigure}%
    \vspace{-1.2\baselineskip}
    \caption{Mel spectrograms of reference and modified results.
    \textbf{(a)} is the input and the lyrics are "Its fleece was white as snow." \textbf{(b)-(d)} are results with the modified attribute.
    }
    \label{fig:control}
\end{figure*}
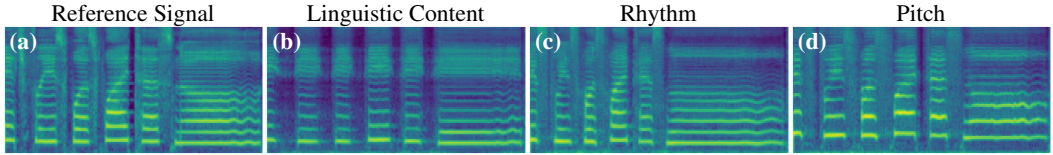

\section{Controllable Singing Voice Synthesis}
Using our model, the linguistic contents, the rhythm, the pitch, and the speaker identity are encoded in interpretable format and hence are easily controllable.
Fig. \ref{fig:control} shows the results of each attribute modified.
More details and samples can be found in Appendix \ref{visualization}, \ref{exist_control}, \ref{user_control} and the demo page\footnote{More samples are available on \href{https://mindslab-ai.github.io/assem-vc/singer/}{https://mindslab-ai.github.io/assem-vc/singer/}}.

\vspace{1pt}\noindent\textbf{Control lyrics.}  
To control the linguistic content of the singing voice, the model first estimates the alignment between the \mel and the original phoneme sequence.
Modified linguistic representation is obtained by matrix multiplication between the estimated alignment and the text encoding from modified phoneme sequence.
Text deletion can be performed by replacing the phonemes with blank tokens and the corresponding pitch values to unvoiced value 0.

\vspace{1pt}\noindent\textbf{Control rhythm.}  
The rhythm is modified by linearly interpolating linguistic and pitch representation.
Since the alignment is estimated, our method enables not only utterance-wise but also per-phoneme control of rhythm using
\textit{argmax} to compute the corresponding temporal indices of linguistic and pitch features of each phoneme in the alignment matrix.

\vspace{1pt}\noindent\textbf{Control pitch.}
The pitch is shifted by multiplying by the powers of $2^{\frac{1}{12}}$, the frequency ratio between adjacent musical notes.
We also observed that switching the pitch to unvoiced value 0 without changing the phoneme sequence generates a whispering voice.

\vspace{1pt}\noindent\textbf{Control speaker identity.}
Switching the speaker embedding changes the singer's voice.

\vspace{1pt}\noindent\textbf{Controlling with voice of the user.}   
We propose a workflow of receiving the user's singing voice as input, encoding each attribute, and converting it into the target singer's singing voice.
We observe that the amount of data in the public singing voice datasets is insufficient to perform a speaker conversion with an arbitrary singer's singing voice as the source, also referred to as any-to-many condition.
Thus, we include a small amount of the user's singing voice in the training set, which is described in Appendix \ref{details}.
We successfully trained the model by adding only two minutes of the author’s singing voice.
In conclusion, we made a perfectly synced duet with the user's singing voice and the target singer's converted singing voice.


\clearpage 

\section*{Broader Impact}
Our system could be applied in entertainment including music productions, video productions, audio book services, and advertisements. However, malicious users could be abused voice synthesis or singing synthesis through crimes such as phishing, fake news, or violating the right of publicity. In addition, since our system needs training on the human singing voice dataset, we applied Children's Song Dataset \cite{csd2020} and NUS-48E \cite{nus48e} with the agreement that it will be used solely for research purposes.

\section*{Acknowledgments}
The authors would like to thank
Sang Hoon Woo and Minho Kim from MINDsLab Inc.,
for providing beneficial feedbacks on the initial draft of this paper.

\bibliographystyle{unsrtnat}
\bibliography{neurips_2021}

\clearpage

\appendix

\captionsetup[table]{skip=5pt}

\section{Experimental Details} \label{details}
\paragraph{Datasets.}
Children's Song Dataset (CSD) \cite{csd2020}, NUS-48E \cite{nus48e}, and the author's singing voice are used for training and evaluation.
We choose two singers from NUS-48E and the author's singing voice as the source.
The other singers in CSD and NUS-48E are used as target singers for conversion.
CSD is an open dataset of English and Korean nursery rhymes sung by professional female singers.
We used 50 English songs, which totals 111 minutes in length.
47 songs are used for training and 3 songs are used for test.
NUS-48E dataset consists of 12 singers' songs with four songs for each singer.
We selected a female \textit{ADIZ} and a male \textit{JLEE} singer to simulate the user's environment, and used only a single song for training and others for testing.
For the other singers in NUS-48E, all songs are used for training, which is a total of 7-11 minutes including silence per each singer.
Author's singing voice was recorded and added during training and testing. 
To simulate a common user environment without professional skills or setups, the author sang one of the popular music, \textit{Fool's Garden}'s \textit{Lemon Tree} into laptop's built-in microphone without any professional equipment.
During the recording process, no effort was made to perfectly match the pitch or the rhythm of the original song.
Detailed statistics of the datasets are shown in Table \ref{tab:dataset}.

\begin{table}[h]
\renewcommand{\arraystretch}{1.3}
\renewcommand{\tabcolsep}{1.59mm}
\caption{%
	Dataset statistics per each speaker used in training scheme. 
	Spk ID. and Len. is short for Speaker ID and Length of dataset, respectively.
	Only the first two letters of the speaker ID of the NUS-48E were displayed.
	Author. is short for author's singing voice.
}
\label{tab:dataset}

\begin{tabular}{|l|c|c|c|c|c|c|c|c|c|c|c|c|c|c|}
\hline
\textbf{Dataset} & \textit{\textbf{CSD}} & \multicolumn{12}{c|}{\textit{\textbf{NUS-48E}}}                                                                                                                       & \textit{\textbf{Author.}} \\ \hline
\textbf{Usage}   & \multicolumn{11}{c|}{\textit{\textbf{Target Singer}}}                                                                                                             & \multicolumn{3}{c|}{\textit{\textbf{User}}}          \\ \hline
\textbf{Spk ID.} & \textit{CSD}          & \textit{JT} & \textit{KE} & \textit{MC} & \textit{MP} & \textit{MP} & \textit{NJ} & \textit{PM} & \textit{SA} & \textit{VK} & \textit{ZH} & \textit{AD} & \textit{JL} & \textit{kwkim}             \\ \hline
\textbf{Gender} & F                     & M           & M           & F           & F           & F           & F           & F           & M           & M           & M           & F           & M           & M                        \\ \hline
\textbf{Len. (min)}    & 105                   & 9.3         & 8.4         & 8.3         & 11        & 7.6         & 9.2         & 10          & 9.0         & 11          & 7.6         & 3.3         & 2.2         & 2.4                      \\ \hline
\end{tabular}
\end{table}
For all datasets, raw audios are resampled to sampling rate 22050Hz and preprocessed to 80-bin log \mel, which is computed from short-time Fourier transform (STFT) with 1024 filter lengths, 1024 window sizes, and 256 hop lengths.
All singing voices are split between 1-12 seconds and used for training with corresponding lyrics.
Additional information such as time-aligned score or phoneme information is not used.

\paragraph{Training Details.}
Our system is trained in three stages: training Cotatron, training decoder with the fixed Cotatron, and ground-truth alignment (GTA) finetuning HiFi-GAN.
To learn the alignment of singing voices with small singing datasets, we transferred Cotatron from pre-trained weights\footnote{We used pre-trained weights of Cotatron contained in pre-trained weights of Assem-VC, which are available on  \href{https://github.com/mindslab-ai/assem-vc}{https://github.com/mindslab-ai/assem-vc}}.
For other training details, we follow the settings of Kim \etal{} \cite{assem-vc}.

\clearpage

\section{Visualization of Decomposition} \label{visualization}
\begin{figure*}[h!]
\setlength{\textfloatsep}{0pt}%

\newcommand{\incaptionimg}[3]{
  \begin{tikzpicture}[every node/.style={inner sep=0,outer sep=0}]
    \draw node[name=micrograph] {\includegraphics[width=\textwidth]{#1}}; 
    \draw  (micrograph.north west)  node[anchor=north west,yshift=-0.25cm,xshift=0.55cm,#3]{\textbf{\small{(#2)}}}; 
  \end{tikzpicture}
}

\captionsetup[subfigure]{labelformat=empty}
    \setlength{\textfloatsep}{0pt}%
    \setlength{\intextsep}{0pt}
    \centering
    \begin{subfigure}[b]{\linewidth}
    \subcaption{\textbf{Reference}}
    \end{subfigure}%
        \vspace{-0.90\baselineskip}
     \begin{subfigure}[b]{\linewidth}
         \incaptionimg{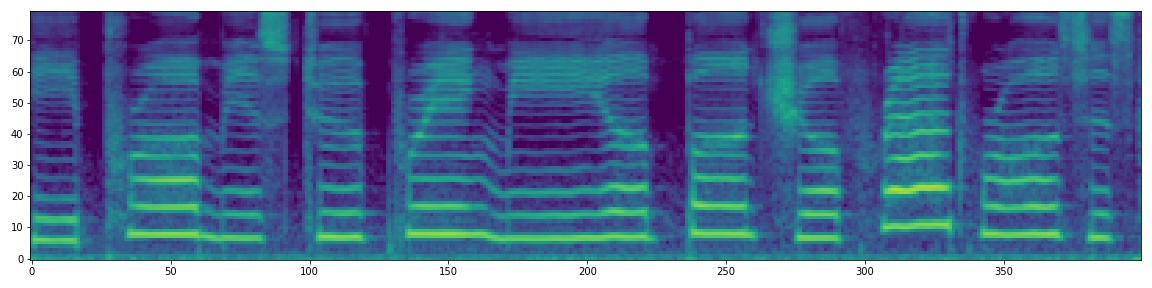}{a}{white}
         \phantomsubcaption\ignorespaces\label{vis_a}
     \end{subfigure}%
    \vspace{-0.90\baselineskip}

    \begin{subfigure}[b]{1.0\linewidth}
    \subcaption{Source Pitch Contour}
    \end{subfigure}%
    \vspace{-0.90\baselineskip}
    \begin{subfigure}[b]{\linewidth}
         \incaptionimg{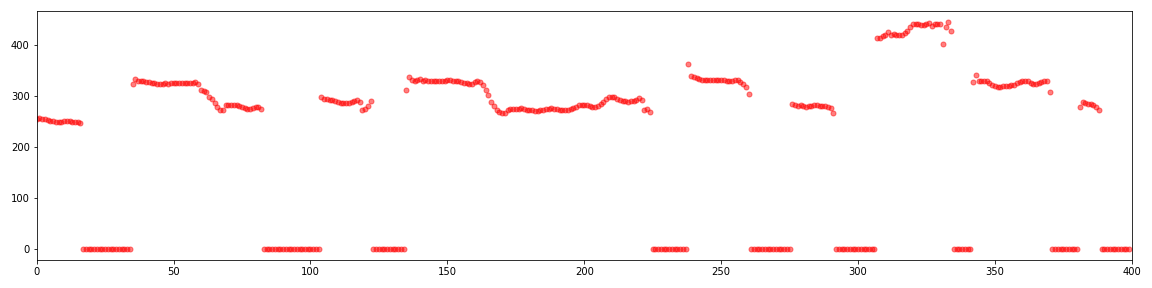}{b}{black}
         \phantomsubcaption\ignorespaces\label{vis_b}
    \end{subfigure}%
    \vspace{-0.90\baselineskip}

    \begin{subfigure}[b]{\linewidth}
    \subcaption{Source Alignment}
    \end{subfigure}%
    \vspace{-0.90\baselineskip}
    \begin{subfigure}[b]{\linewidth}
         \incaptionimg{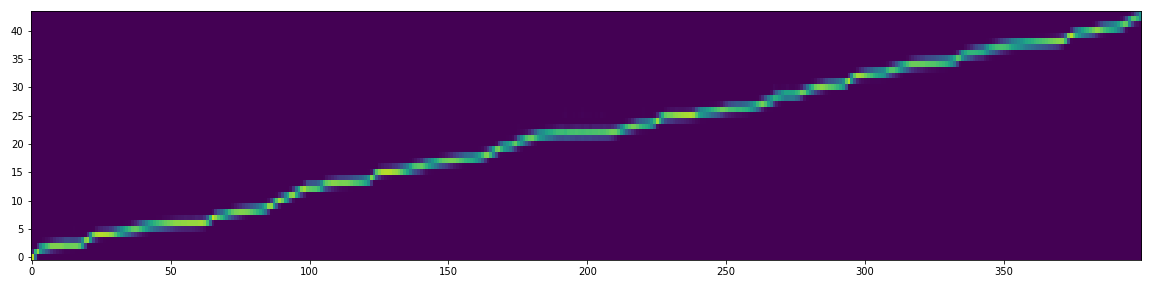}{c}{white}
         \phantomsubcaption\ignorespaces\label{vis_c}
    \end{subfigure}%
    \vspace{-0.90\baselineskip}

    \begin{subfigure}[b]{\linewidth}
    \subcaption{Decoder Prediction}
    \end{subfigure}%
    \vspace{-0.90\baselineskip}
    \begin{subfigure}[b]{\linewidth}
         \incaptionimg{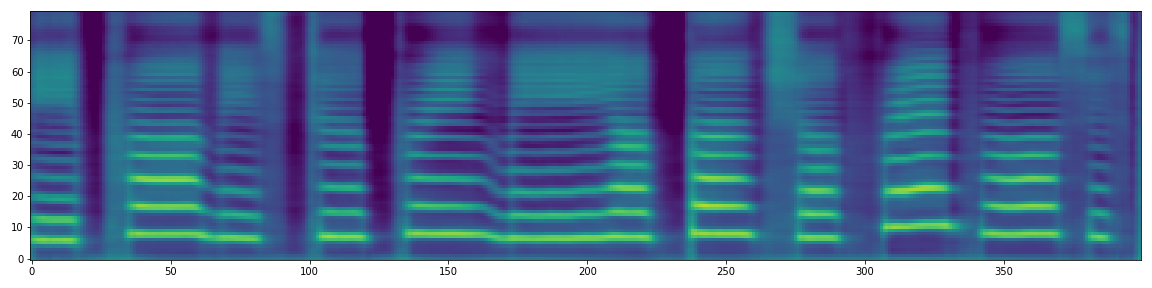}{d}{white}
         \phantomsubcaption\ignorespaces\label{vis_d}
    \end{subfigure}%
    \vspace{-0.90\baselineskip}

    \begin{subfigure}[b]{\linewidth}
    \subcaption{Vocoder Output}
    \end{subfigure}%
    \vspace{-0.90\baselineskip}
    \begin{subfigure}[b]{\linewidth}
         \incaptionimg{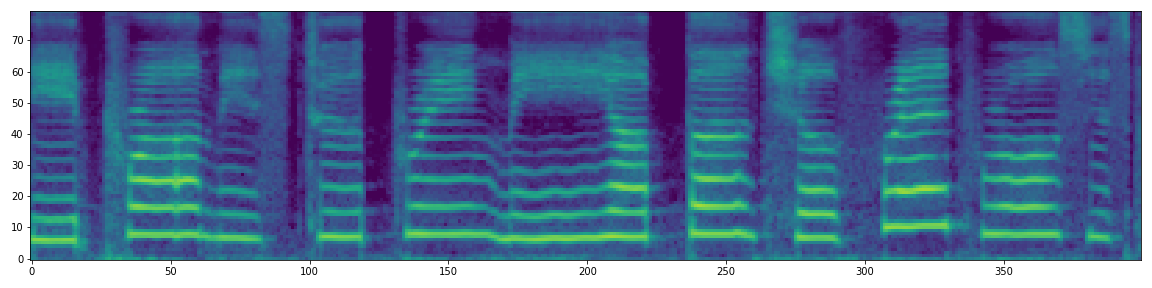}{e}{white}
         \phantomsubcaption\ignorespaces\label{vis_e}
    \end{subfigure}%
    \vspace{-0.90\baselineskip}

    \vspace{-1.2\baselineskip}
    \caption{Visualization of singing voice decomposition. Our model estimates the pitch contour \textbf{(b)} and the alignment \textbf{(c)} from reference audio \textbf{(a)}. The input lyrics are "He promised to bring me a bunch of red roses." The decoder reconstructs the \mel \textbf{(d)}, which in turn, is converted to raw audio by GTA finetuned HiFi-GAN. \textbf{(e)} is the \mel of the output raw audio.
    }
    \label{fig:visualize}
\end{figure*}
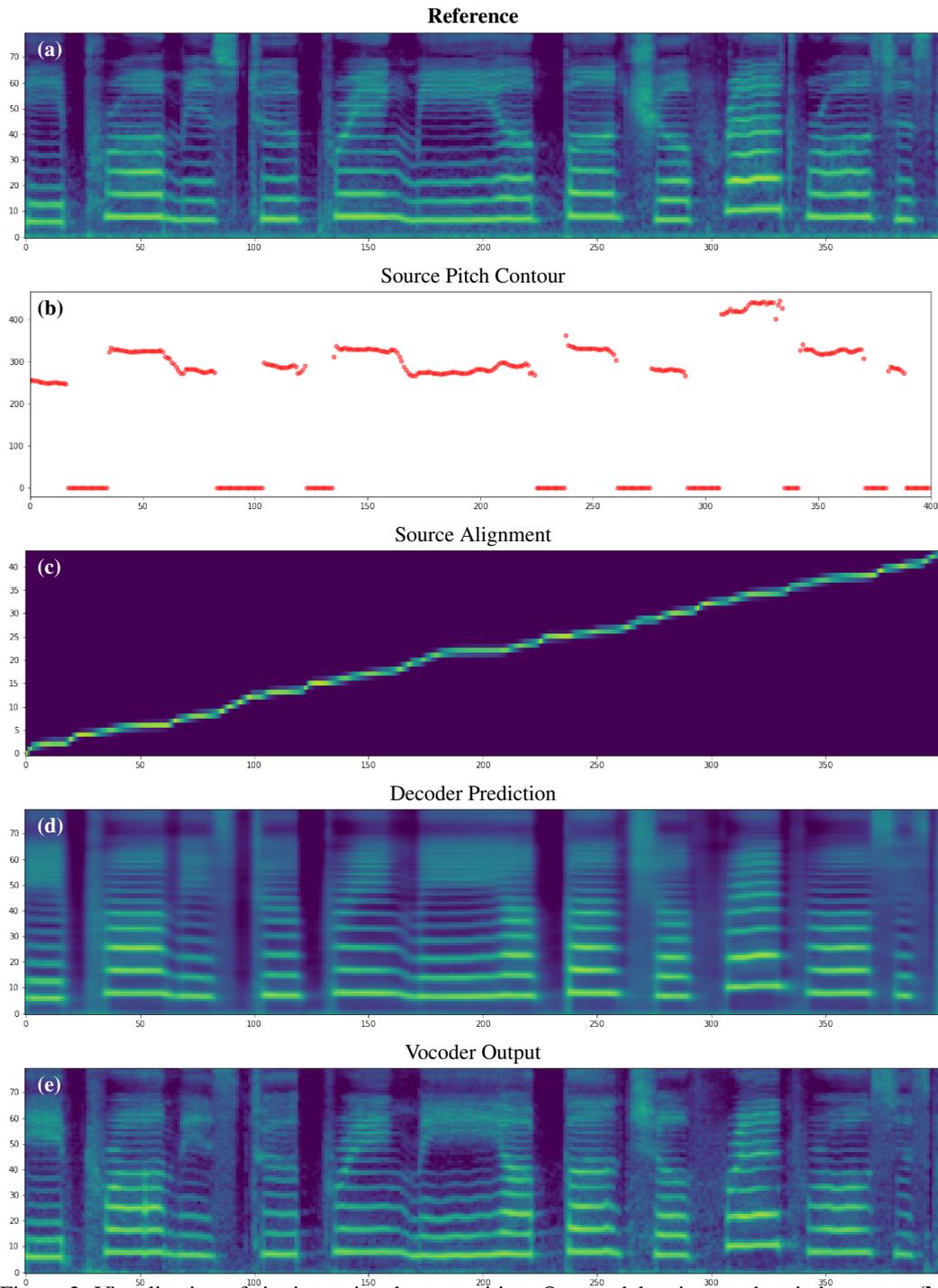

\clearpage

\section{Controlling each attribute of an existing singing voice.} \label{exist_control}
\subsection{Control lyrics}

\begin{figure*}[h!]
\setlength{\textfloatsep}{0pt}%

\newcommand{\incaptionimg}[3]{
  \begin{tikzpicture}[every node/.style={inner sep=0,outer sep=0}]
    \draw node[name=micrograph] {\includegraphics[width=\textwidth]{#1}}; 
    \draw  (micrograph.north west)  node[anchor=north west,yshift=-0.25cm,xshift=0.45cm,#3]{\textbf{\small{(#2)}}}; 
  \end{tikzpicture}
}

\captionsetup[subfigure]{labelformat=empty}
    \setlength{\textfloatsep}{0pt}%
    \setlength{\intextsep}{0pt}
    \centering
    \begin{subfigure}[b]{\linewidth}
    \subcaption{\textbf{Reference} \{OW\} \{D IH R\} \{W AH T\} \{K AE N\} \{DH AH\} \{M AE T ER\} \{B IY\}.}
    \end{subfigure}%
        \vspace{-0.90\baselineskip}
     \begin{subfigure}[b]{\linewidth}
         \incaptionimg{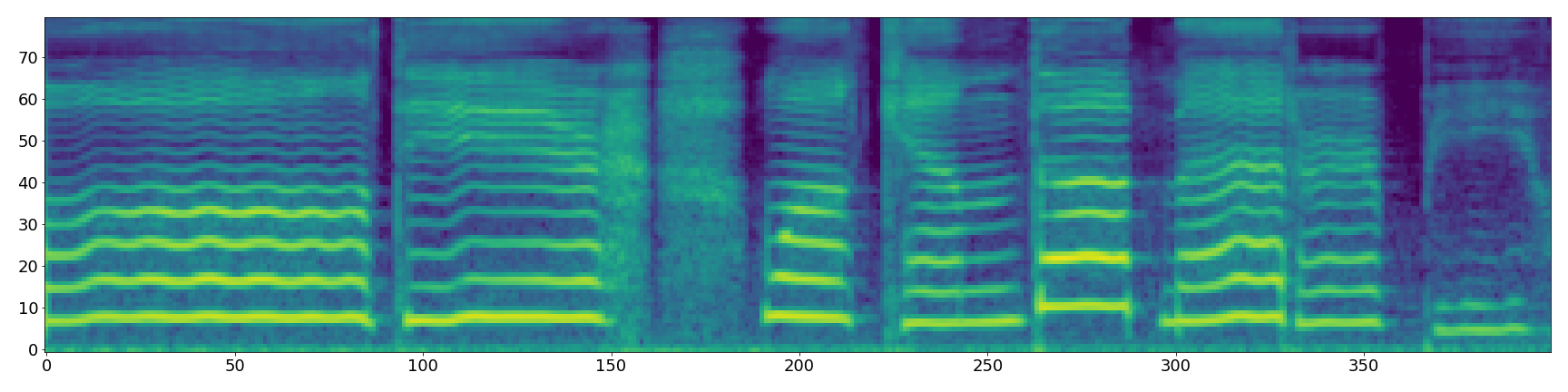}{a}{white}
         \phantomsubcaption\ignorespaces\label{lyric_a}
     \end{subfigure}%
    \vspace{-0.90\baselineskip}

    \begin{subfigure}[b]{\linewidth}
    \subcaption{{\color{red} \{UH\} \{DH ER R\} \{M AW S\}} \{K AE N\} {\color{red} \{N AA\} \{B AA DH ER\} \{M IY\}}.}
    \end{subfigure}%
    \vspace{-0.90\baselineskip}
    \begin{subfigure}[b]{\linewidth}
         \incaptionimg{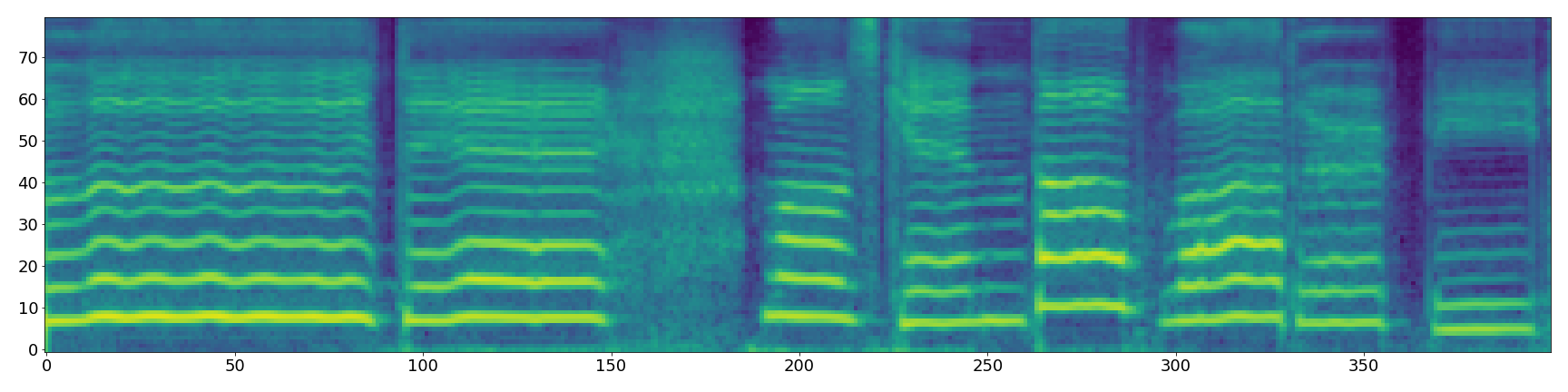}{b}{white}
         \phantomsubcaption\ignorespaces\label{lyric_b}
    \end{subfigure}%
    \vspace{-0.90\baselineskip}

    \begin{subfigure}[b]{\linewidth}
    \subcaption{{\color{red} \{AH\} \{AH AH AH\} \{AH AH AH\} \{AH AH AH\} \{AH AH\} \{AH AH AH AH\} \{AH AH\}}.}
    \end{subfigure}%
    \vspace{-0.90\baselineskip}
    \begin{subfigure}[b]{\linewidth}
         \incaptionimg{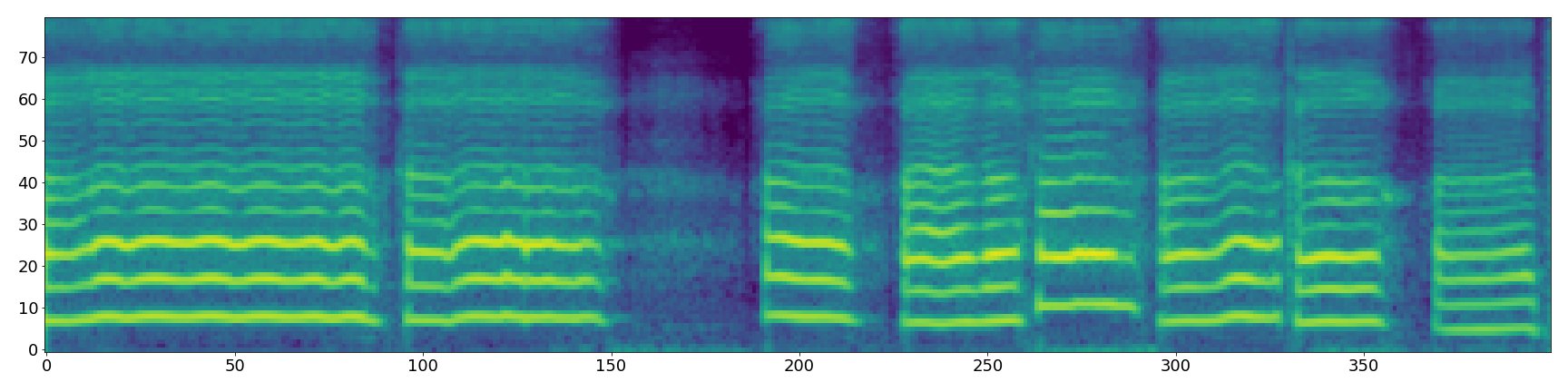}{c}{white}
         \phantomsubcaption\ignorespaces\label{lyric_c}
    \end{subfigure}%
    \vspace{-0.90\baselineskip}

    \begin{subfigure}[b]{\linewidth}
    \subcaption{{\color{red} BLANK}}
    \end{subfigure}%
    \vspace{-0.90\baselineskip}
    \begin{subfigure}[b]{\linewidth}
         \incaptionimg{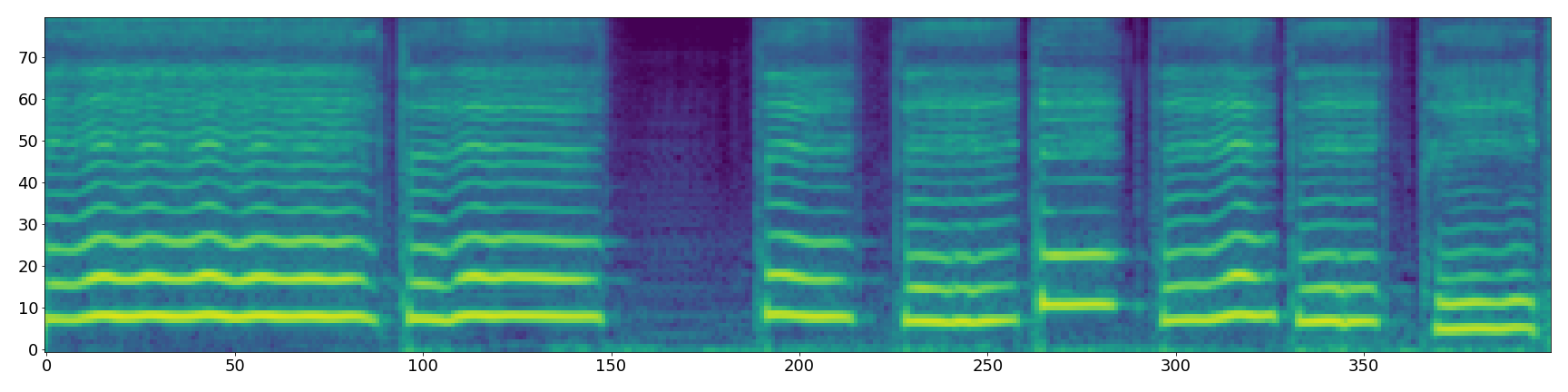}{d}{white}
         \phantomsubcaption\ignorespaces\label{lyric_d}
    \end{subfigure}%
    \vspace{-0.90\baselineskip}

    \begin{subfigure}[b]{\linewidth}
    \subcaption{\{OW\} \{D IH R\} {\color{red} \deleted{\{W AH T\}}} \{K AE N\} \{DH AH\} \{M AE T ER\} \{B IY\}.}
    \end{subfigure}%
    \vspace{-0.90\baselineskip}
    \begin{subfigure}[b]{\linewidth}
         \incaptionimg{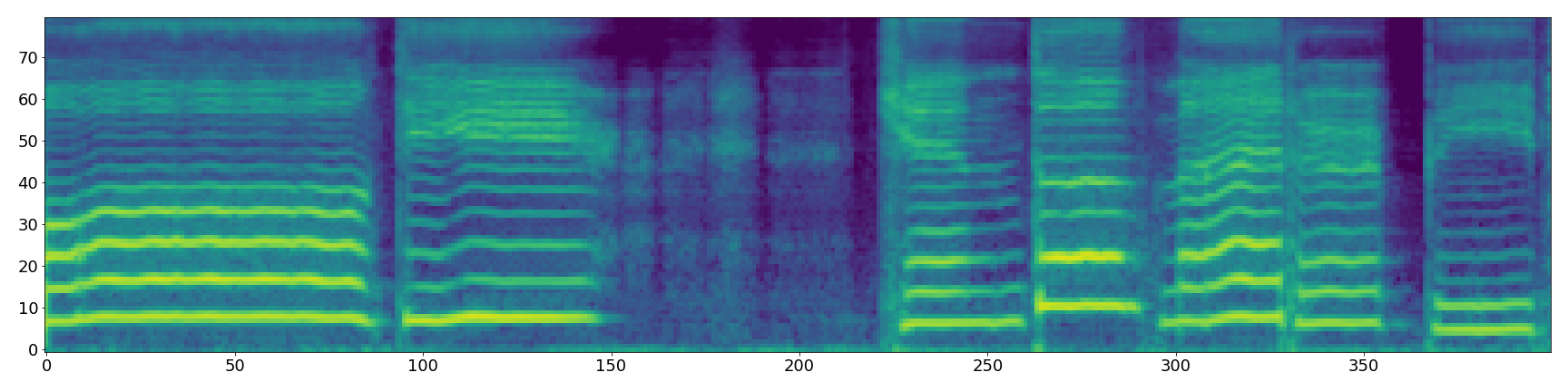}{e}{white}
         \phantomsubcaption\ignorespaces\label{lyric_e}
    \end{subfigure}%
    \vspace{-0.90\baselineskip}

    \vspace{-1.2\baselineskip}
    \caption{Mel spectrograms of the results with modified lyrics.
    \textbf{(a)} is the input and the lyrics are "Oh dear what can the matter be." 
    \textbf{(b)-(e)} are samples with the modified lyrics.
    The texts in red denote the changes in the lyrics.
    BLANK indicates that all phonemes were replaced with blank tokens without changing the pitches.
    Crossed out texts represent the phonemes replaced with blank tokens and the corresponding pitches with 0.
    }
    \label{fig:control_lyric}
\end{figure*}
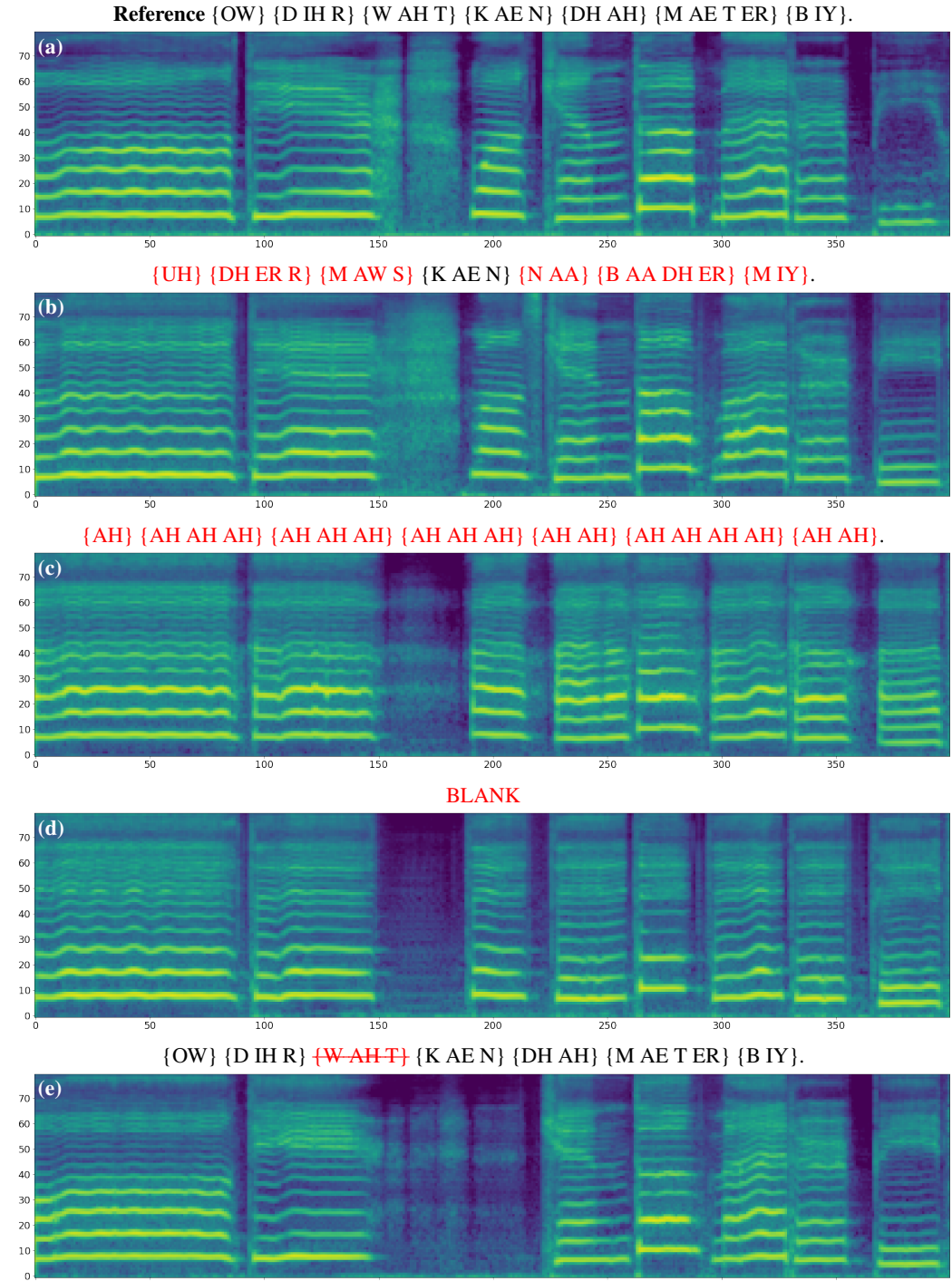

\clearpage
\subsection{Control rhythm}
\begin{figure*}[h!]
\setlength{\textfloatsep}{0pt}%

\newcommand{\incaptionimg}[3]{
  \begin{tikzpicture}[every node/.style={inner sep=0,outer sep=0}]
    \draw node[name=micrograph] {\includegraphics[width=\textwidth]{#1}}; 
    \draw  (micrograph.north west)  node[anchor=north west,yshift=-0.25cm,xshift=0.45cm,#3]{\textbf{\small{(#2)}}}; 
  \end{tikzpicture}
}

\captionsetup[subfigure]{labelformat=empty}
    \setlength{\textfloatsep}{0pt}%
    \setlength{\intextsep}{0pt}
    \centering
    \begin{subfigure}[b]{\linewidth}
    \subcaption{\textbf{Reference} \{IH T S\} \{F L IY S\} \{W AA Z\} \{W AY T\} \{AE Z\} \{S N OW\}.}
    \end{subfigure}%
        \vspace{-0.90\baselineskip}
     \begin{subfigure}[b]{\linewidth}
         \incaptionimg{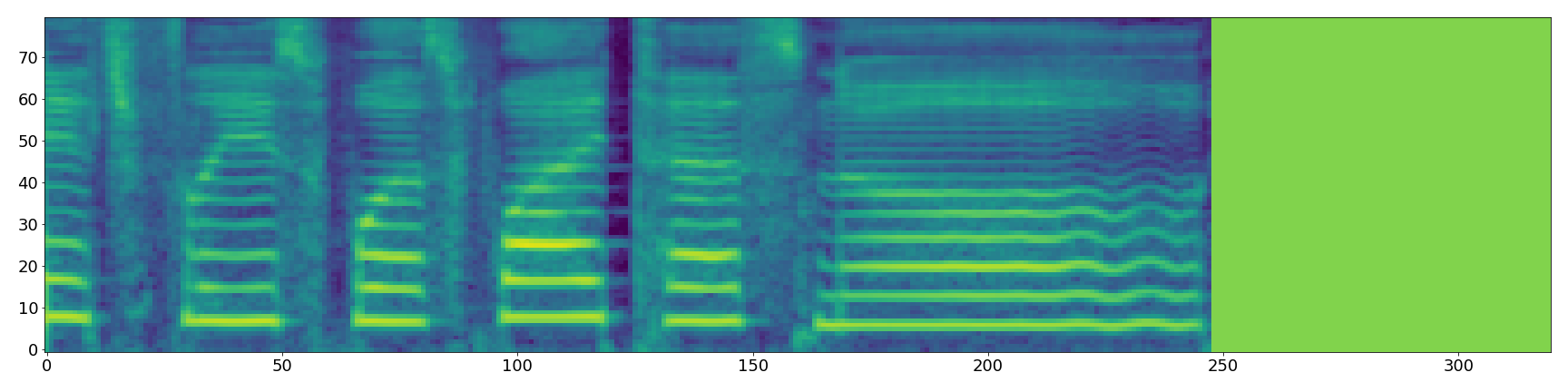}{a}{white}
         \phantomsubcaption\ignorespaces\label{rhythm_a}
     \end{subfigure}%
    \vspace{-0.90\baselineskip}

    \begin{subfigure}[b]{\linewidth}
    \subcaption{\{IH T S\} \{F L IY S\} \{W AA Z\} \{W AY T\} \{AE Z\} \{S N {\color{red} OW}\}. {\color{red} $\times$0.5 }}
    \end{subfigure}%
    \vspace{-0.90\baselineskip}
    \begin{subfigure}[b]{\linewidth}
         \incaptionimg{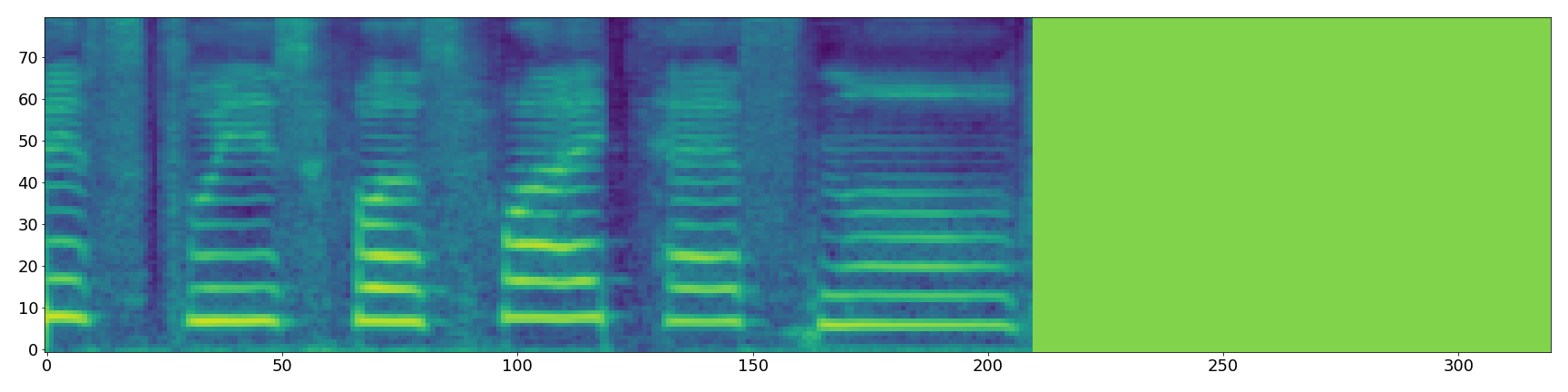}{b}{white}
         \phantomsubcaption\ignorespaces\label{rhythm_b}
    \end{subfigure}%
    \vspace{-0.90\baselineskip}

    \begin{subfigure}[b]{\linewidth}
    \subcaption{\{IH T S\} \{F L IY S\} \{W AA Z\} \{W AY T\} \{AE Z\} \{S N {\color{red} OW}\}. {\color{red} $\times$2.5 }}
    \end{subfigure}%
    \vspace{-0.90\baselineskip}
    \begin{subfigure}[b]{\linewidth}
         \incaptionimg{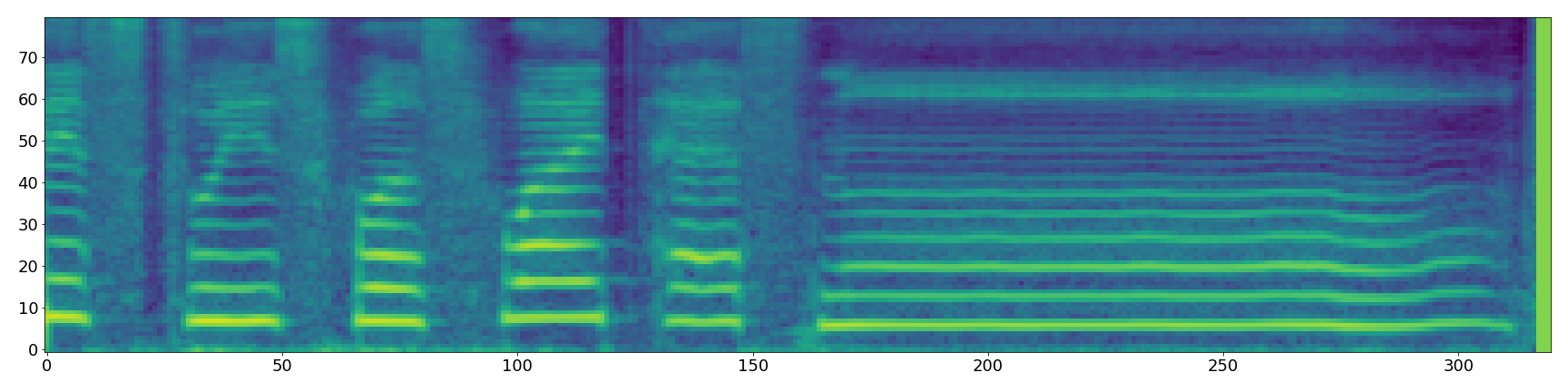}{c}{white}
         \phantomsubcaption\ignorespaces\label{rhythm_c}
    \end{subfigure}%
    \vspace{-0.90\baselineskip}

    \begin{subfigure}[b]{\linewidth}
    \subcaption{\{IH T S\} \{F L IY S\} \{W AA Z\} \{W AY T\} \{AE Z\} \{{\color{red} S} N OW\}. {\color{red} $\times$5 }}
    \end{subfigure}%
    \vspace{-0.90\baselineskip}
    \begin{subfigure}[b]{\linewidth}
         \incaptionimg{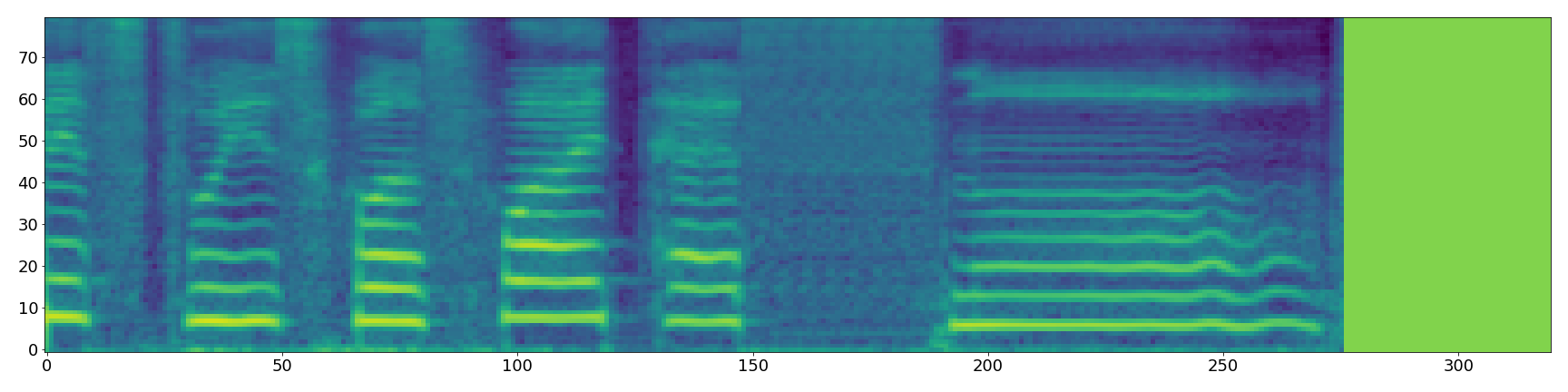}{d}{white}
         \phantomsubcaption\ignorespaces\label{rhythm_d}
    \end{subfigure}%
    \vspace{-0.90\baselineskip}

    \begin{subfigure}[b]{\linewidth}
    \subcaption{\{IH T S\} \{F L IY S\} \{W AA Z\} \{W AY T\} \{{\color{red} AE} Z\} \{S N OW\}. {\color{red} $\times$5 }}
    \end{subfigure}%
    \vspace{-0.90\baselineskip}
    \begin{subfigure}[b]{\linewidth}
         \incaptionimg{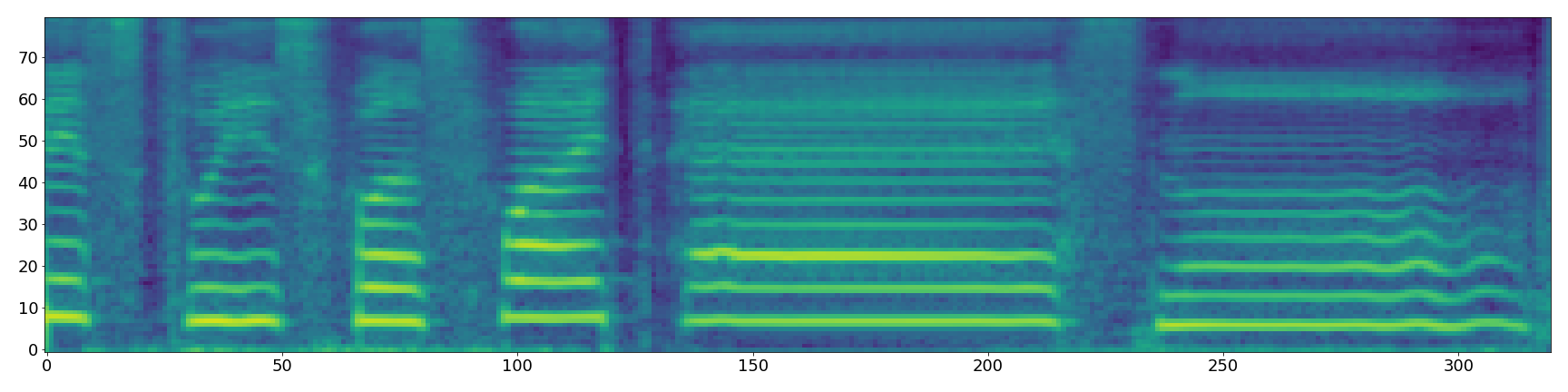}{e}{white}
         \phantomsubcaption\ignorespaces\label{rhythm_e}
    \end{subfigure}%
    \vspace{-0.90\baselineskip}

    \vspace{-1.2\baselineskip}
    \caption{Mel spectrograms of the results with the modified rhythm.
    \textbf{(a)} is the input \mel and the lyrics are "Its fleece was white as snow." 
    \textbf{(b)-(e)} are samples with the modified rhythms.
    The texts in red indicate the phonemes with modified duration, and the numbers at the end denote the resampling factor.
    }
    \label{fig:control_rhythm}
\end{figure*}
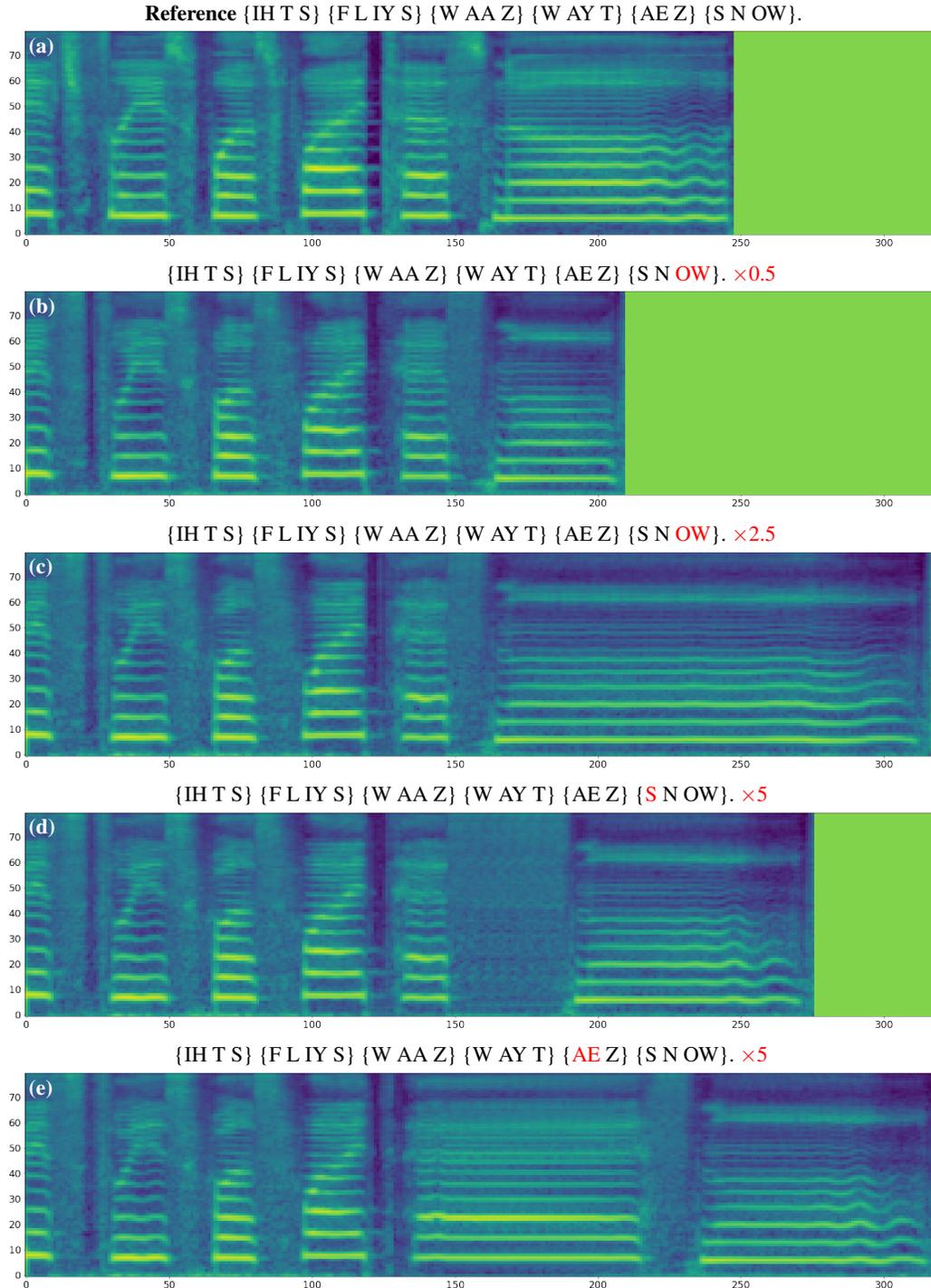

\clearpage

\subsection{Control pitch}
\begin{figure*}[h!]
\setlength{\textfloatsep}{0pt}%

\newcommand{\incaptionimg}[3]{
  \begin{tikzpicture}[every node/.style={inner sep=0,outer sep=0}]
    \draw node[name=micrograph] {\includegraphics[width=\textwidth]{#1}}; 
    \draw  (micrograph.north west)  node[anchor=north west,yshift=-0.125cm,xshift=0.25cm,#3]{\textbf{\small{(#2)}}}; 
  \end{tikzpicture}
}

\captionsetup[subfigure]{labelformat=empty}
    \setlength{\textfloatsep}{0pt}%
    \setlength{\intextsep}{0pt}
    \centering
    \begin{subfigure}[b]{0.3333\linewidth}
    \subcaption{$-6$}
    \end{subfigure}%
    \begin{subfigure}[b]{0.3333\linewidth}
    \subcaption{$-5$}
    \end{subfigure}%
    \begin{subfigure}[b]{0.3333\linewidth}
    \subcaption{$-4$}
    \end{subfigure}%
    \vspace{-0.90\baselineskip}

    \begin{subfigure}[b]{0.3333\linewidth}
         \incaptionimg{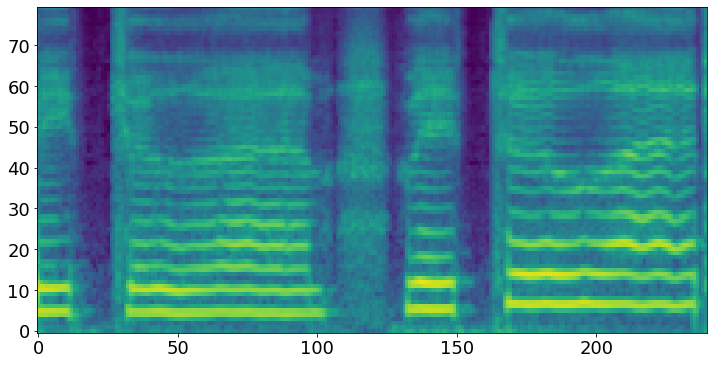}{a}{white}
         \phantomsubcaption\ignorespaces\label{pitch_a}
    \end{subfigure}%
    \begin{subfigure}[b]{0.3333\linewidth}
         \incaptionimg{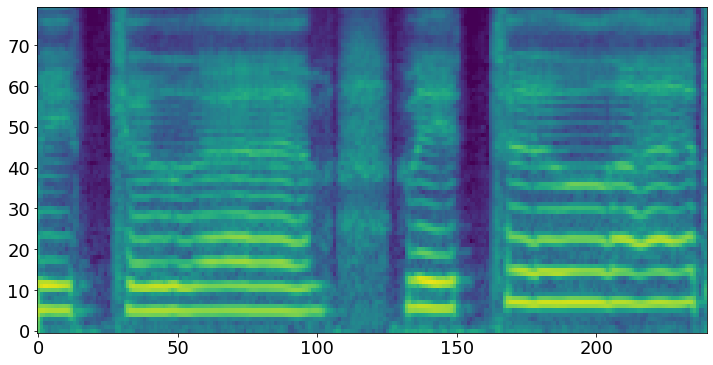}{b}{white}
         \phantomsubcaption\ignorespaces\label{pitch_b}
    \end{subfigure}%
    \begin{subfigure}[b]{0.3333\linewidth}
         \incaptionimg{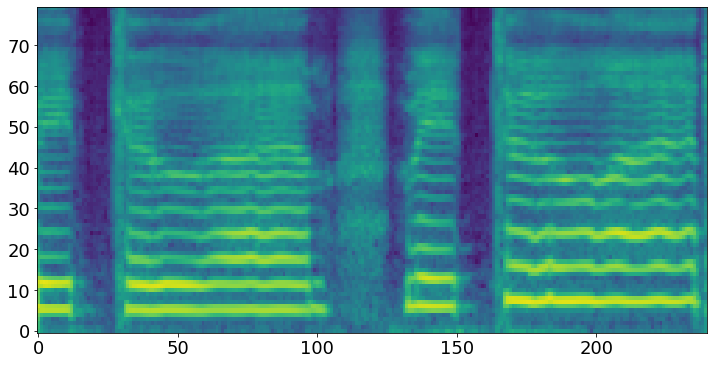}{c}{white}
         \phantomsubcaption\ignorespaces\label{pitch_c}
    \end{subfigure}%
    \vspace{-0.90\baselineskip}

    \begin{subfigure}[b]{0.3333\linewidth}
    \subcaption{$-3$}
    \end{subfigure}%
    \begin{subfigure}[b]{0.3333\linewidth}
    \subcaption{$-2$}
    \end{subfigure}%
    \begin{subfigure}[b]{0.3333\linewidth}
    \subcaption{$-1$}
    \end{subfigure}%
    \vspace{-0.90\baselineskip}

    \begin{subfigure}[b]{0.3333\linewidth}
         \incaptionimg{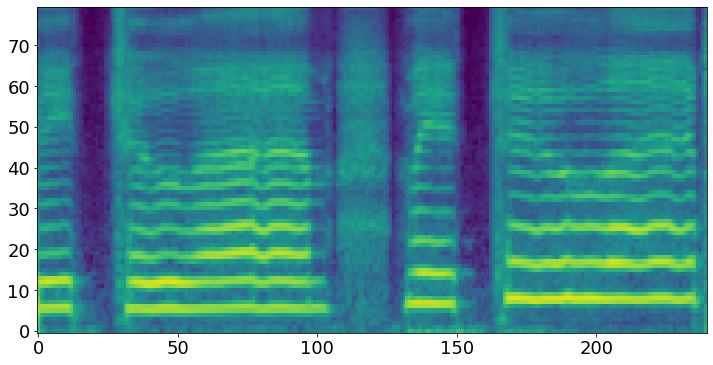}{d}{white}
         \phantomsubcaption\ignorespaces\label{pitch_d}
    \end{subfigure}%
    \begin{subfigure}[b]{0.3333\linewidth}
         \incaptionimg{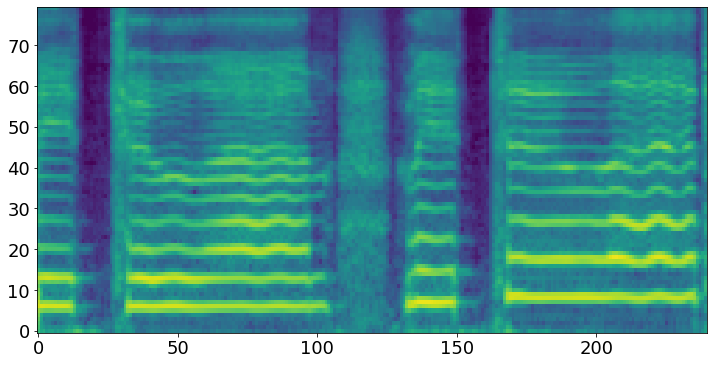}{e}{white}
         \phantomsubcaption\ignorespaces\label{pitch_e}
    \end{subfigure}%
    \begin{subfigure}[b]{0.3333\linewidth}
         \incaptionimg{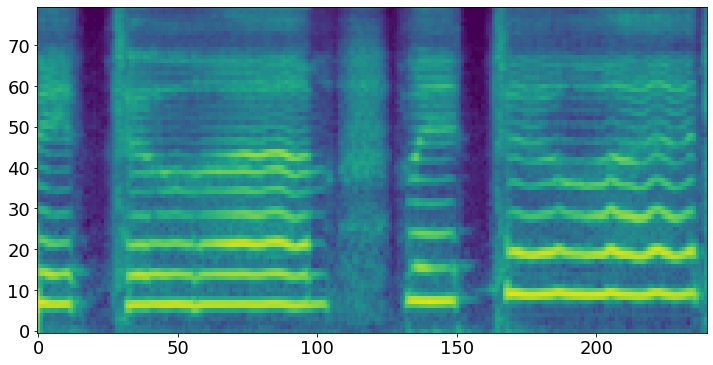}{f}{white}
         \phantomsubcaption\ignorespaces\label{pitch_f}
    \end{subfigure}%
    \vspace{-0.90\baselineskip}

    \begin{subfigure}[b]{0.3333\linewidth}
    \subcaption{\textbf{Reference}}
    \end{subfigure}%
    \begin{subfigure}[b]{0.3333\linewidth}
    \subcaption{$+1$}
    \end{subfigure}%
    \begin{subfigure}[b]{0.3333\linewidth}
    \subcaption{$+2$}
    \end{subfigure}%
    \vspace{-0.90\baselineskip}

    \begin{subfigure}[b]{0.3333\linewidth}
         \incaptionimg{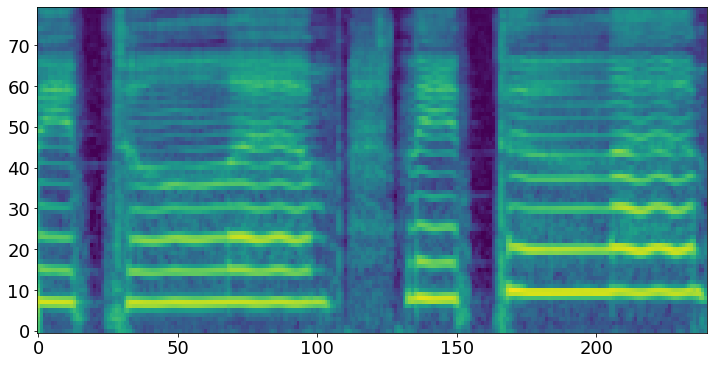}{g}{white}
         \phantomsubcaption\ignorespaces\label{pitch_g}
    \end{subfigure}%
    \begin{subfigure}[b]{0.3333\linewidth}
         \incaptionimg{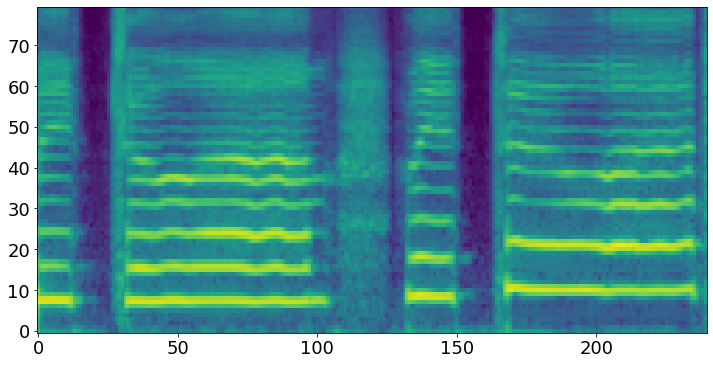}{h}{white}
         \phantomsubcaption\ignorespaces\label{pitch_h}
    \end{subfigure}%
    \begin{subfigure}[b]{0.3333\linewidth}
         \incaptionimg{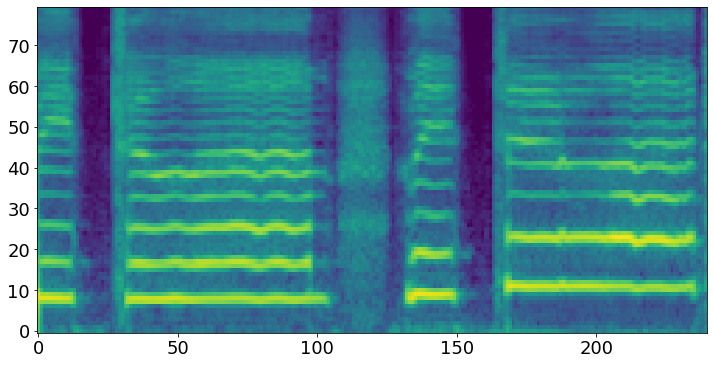}{i}{white}
         \phantomsubcaption\ignorespaces\label{pitch_i}
    \end{subfigure}%
    \vspace{-0.90\baselineskip}

    \begin{subfigure}[b]{0.3333\linewidth}
    \subcaption{$+3$}
    \end{subfigure}%
    \begin{subfigure}[b]{0.3333\linewidth}
    \subcaption{$+4$}
    \end{subfigure}%
    \begin{subfigure}[b]{0.3333\linewidth}
    \subcaption{$+5$}
    \end{subfigure}%
    \vspace{-0.90\baselineskip}

    \begin{subfigure}[b]{0.3333\linewidth}
         \incaptionimg{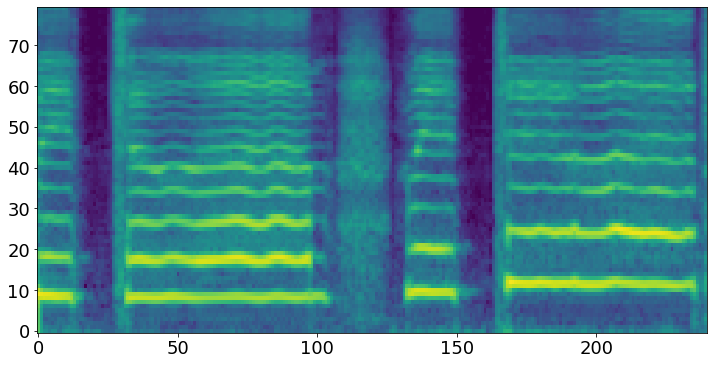}{j}{white}
         \phantomsubcaption\ignorespaces\label{pitch_j}
    \end{subfigure}%
    \begin{subfigure}[b]{0.3333\linewidth}
         \incaptionimg{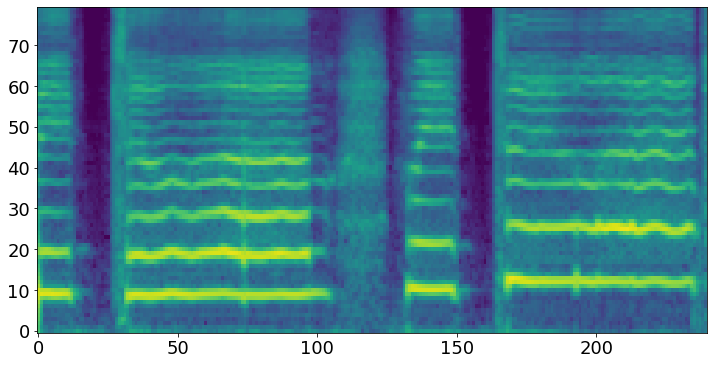}{k}{white}
         \phantomsubcaption\ignorespaces\label{pitch_k}
    \end{subfigure}%
    \begin{subfigure}[b]{0.3333\linewidth}
         \incaptionimg{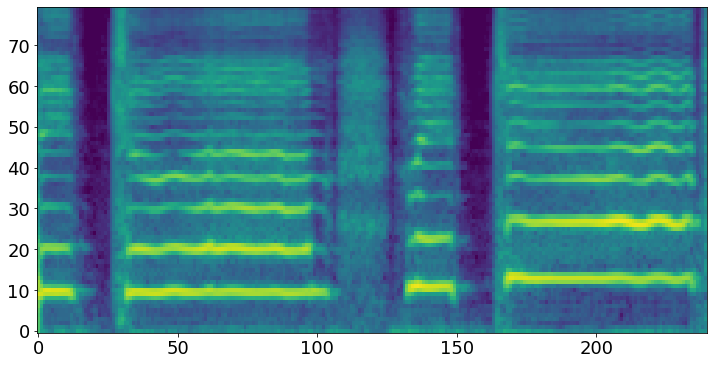}{l}{white}
         \phantomsubcaption\ignorespaces\label{pitch_l}
    \end{subfigure}%

    \vspace{-1.2\baselineskip}
    \caption{Mel spectrograms of the results with the shifted pitch.
    \textbf{(g)} is the input \mel and the lyrics are "Little lamb, little lamb." 
    \textbf{(a)-(f)} and \textbf{(h)-(l)} are samples with the shifted pitches.
    }
    \label{fig:control_pitch}
\end{figure*}
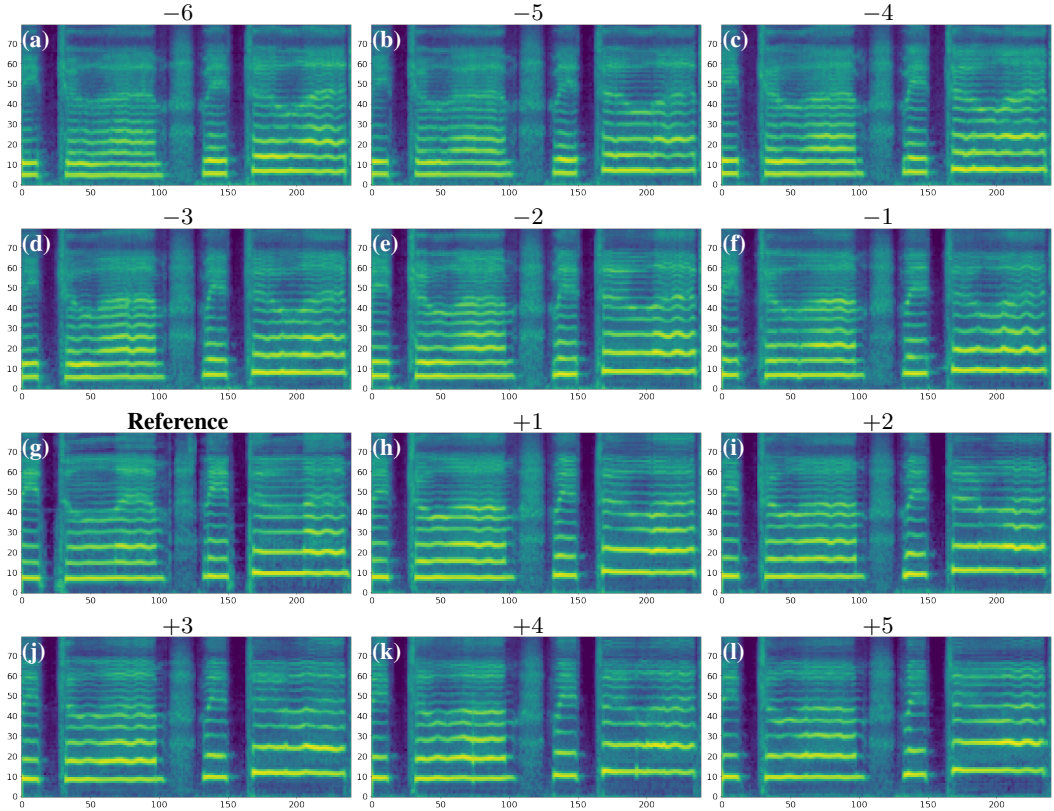

\begin{figure*}[h!]
\setlength{\textfloatsep}{0pt}%

\newcommand{\incaptionimg}[3]{
  \begin{tikzpicture}[every node/.style={inner sep=0,outer sep=0}]
    \draw node[name=micrograph] {\includegraphics[width=\textwidth]{#1}}; 
    \draw  (micrograph.north west)  node[anchor=north west,yshift=-0.25cm,xshift=0.45cm,#3]{\textbf{\small{(#2)}}}; 
  \end{tikzpicture}
}
\captionsetup[subfigure]{labelformat=empty}
    \setlength{\textfloatsep}{0pt}%
    \setlength{\intextsep}{0pt}
    \centering
    \begin{subfigure}[b]{\linewidth}
    \subcaption{\textbf{Reference} \{OW\} \{D IH R\} \{W AH T\} \{K AE N\} \{DH AH\} \{M AE T ER\} \{B IY\}.}
    \end{subfigure}%
        \vspace{-0.90\baselineskip}
     \begin{subfigure}[b]{\linewidth}
         \incaptionimg{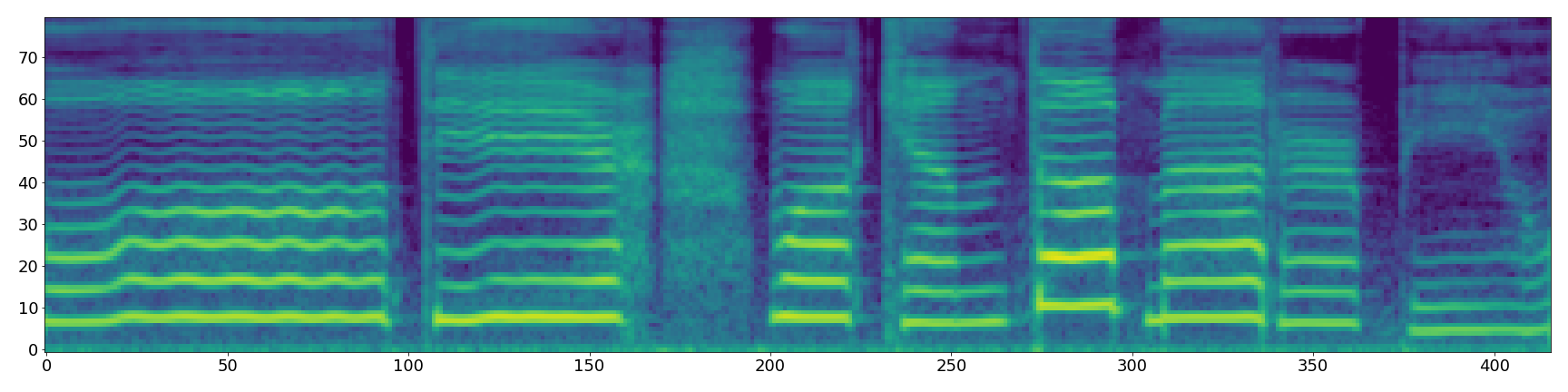}{a}{white}
         \phantomsubcaption\ignorespaces\label{pd_a}
     \end{subfigure}%
         \vspace{-0.90\baselineskip}
    \begin{subfigure}[b]{\linewidth}
    \subcaption{\{OW\} \{D IH R\} \{W AH T\} {\color{red} \deleted{\{K AE N\} \{DH AH\} \{M AE T ER\} \{B IY\}}}.}
    \end{subfigure}%
        \vspace{-0.90\baselineskip}
     \begin{subfigure}[b]{\linewidth}
         \incaptionimg{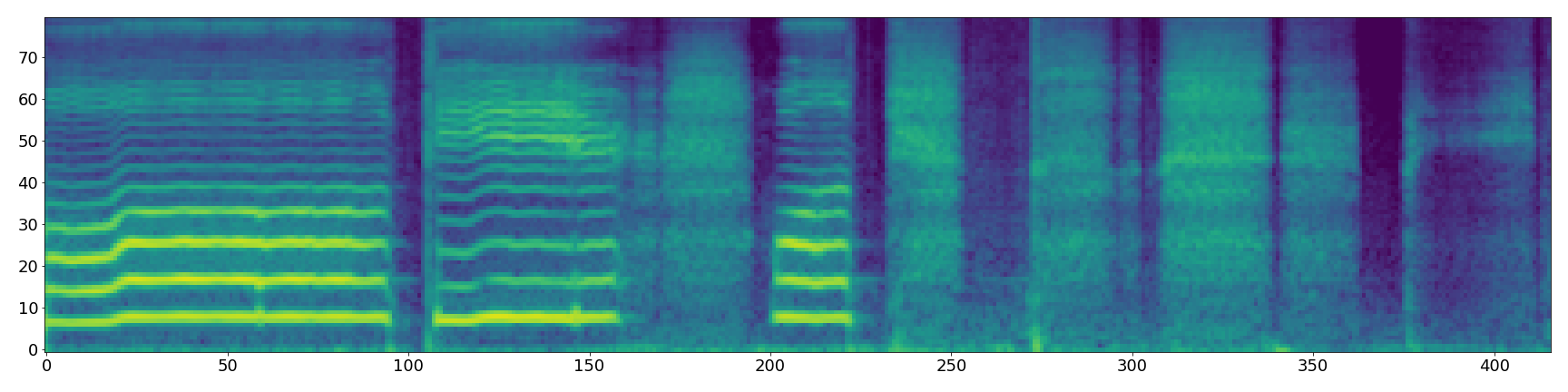}{b}{white}
         \phantomsubcaption\ignorespaces\label{pd_b}
     \end{subfigure}%
         \vspace{-0.90\baselineskip}

    \vspace{-1.2\baselineskip}
    \caption{Mel spectrograms of the results with the deleted pitch.
    \textbf{(a)} is the input \mel and the lyrics are "Oh dear what can the matter be."
    \textbf{(b)} is sample with the deleted pitches.
    The pitches corresponding to crossed-out texts are replaced to unvoiced value 0. The pitch-deleted parts sound like whispers.
    }

    \label{fig:pitch_deletion}
\end{figure*}
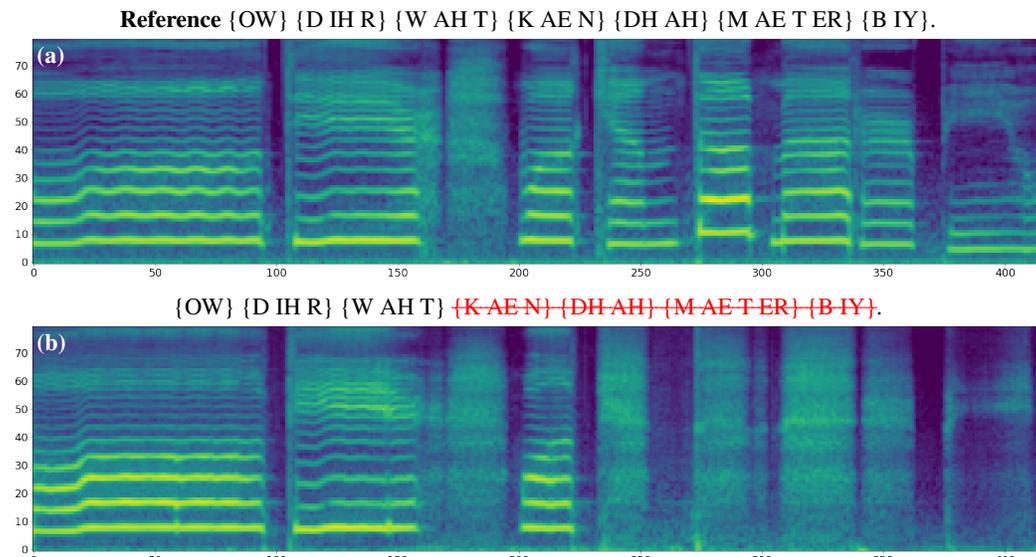

\clearpage

\subsection{Control speaker identity}
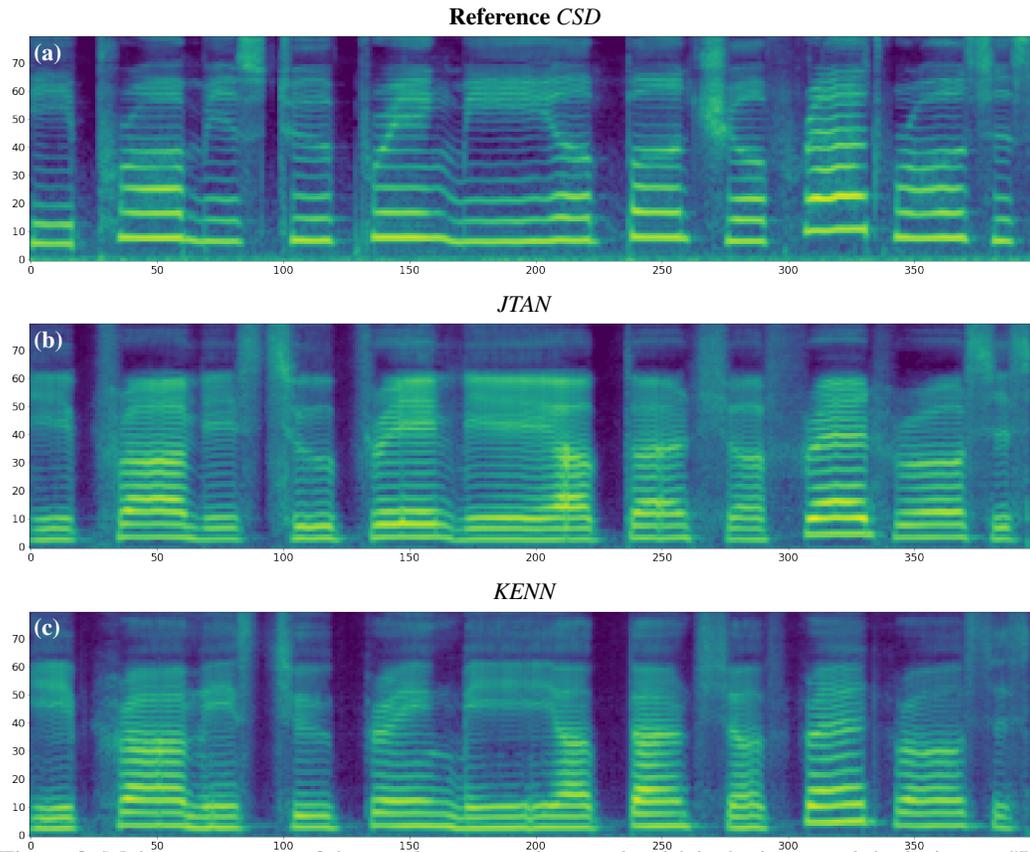
\begin{figure*}[h!]
\setlength{\textfloatsep}{0pt}%

\newcommand{\incaptionimg}[3]{
  \begin{tikzpicture}[every node/.style={inner sep=0,outer sep=0}]
    \draw node[name=micrograph] {\includegraphics[width=\textwidth]{#1}}; 
    \draw  (micrograph.north west)  node[anchor=north west,yshift=-0.25cm,xshift=0.45cm,#3]{\textbf{\small{(#2)}}}; 
  \end{tikzpicture}
}

\captionsetup[subfigure]{labelformat=empty}
    \setlength{\textfloatsep}{0pt}%
    \setlength{\intextsep}{0pt}
    \centering
    \begin{subfigure}[b]{\linewidth}
    \subcaption{\textbf{Reference} \textit{CSD}}
    \end{subfigure}%
        \vspace{-0.90\baselineskip}
     \begin{subfigure}[b]{\linewidth}
         \incaptionimg{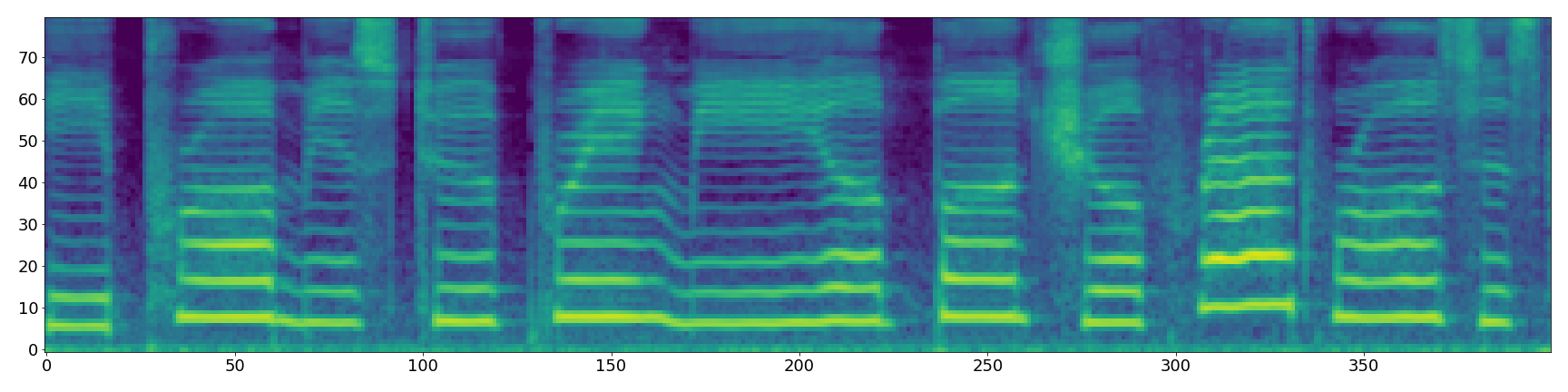}{a}{white}
         \phantomsubcaption\ignorespaces\label{speaker_a}
     \end{subfigure}%
    \vspace{-0.90\baselineskip}

    \begin{subfigure}[b]{\linewidth}
    \subcaption{\textit{JTAN}}
    \end{subfigure}%
    \vspace{-0.90\baselineskip}
    \begin{subfigure}[b]{\linewidth}
         \incaptionimg{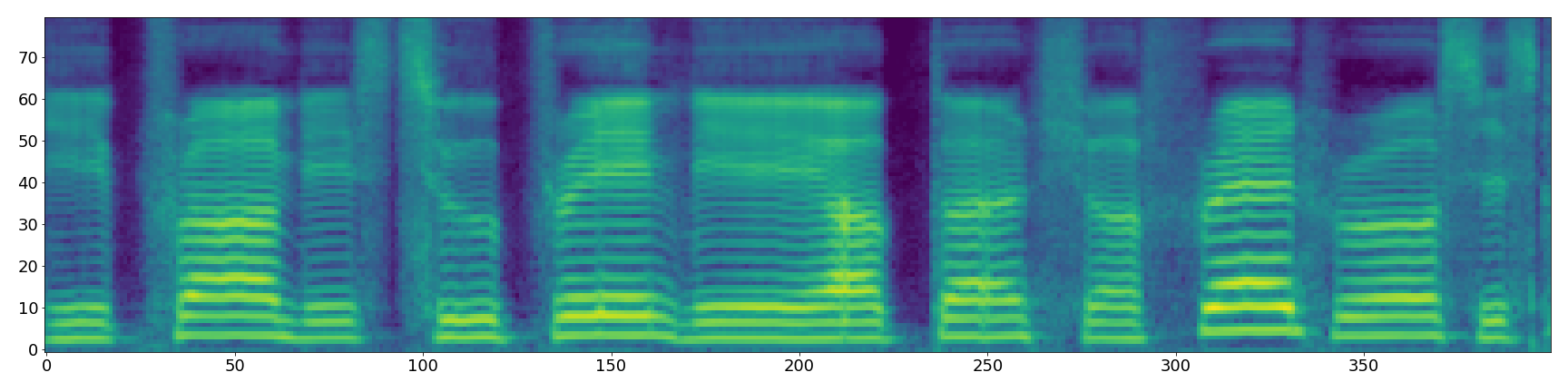}{b}{white}
         \phantomsubcaption\ignorespaces\label{speaker_b}
    \end{subfigure}%
    \vspace{-0.90\baselineskip}

    \begin{subfigure}[b]{\linewidth}
    \subcaption{\textit{KENN}}
    \end{subfigure}%
    \vspace{-0.90\baselineskip}
    \begin{subfigure}[b]{\linewidth}
         \incaptionimg{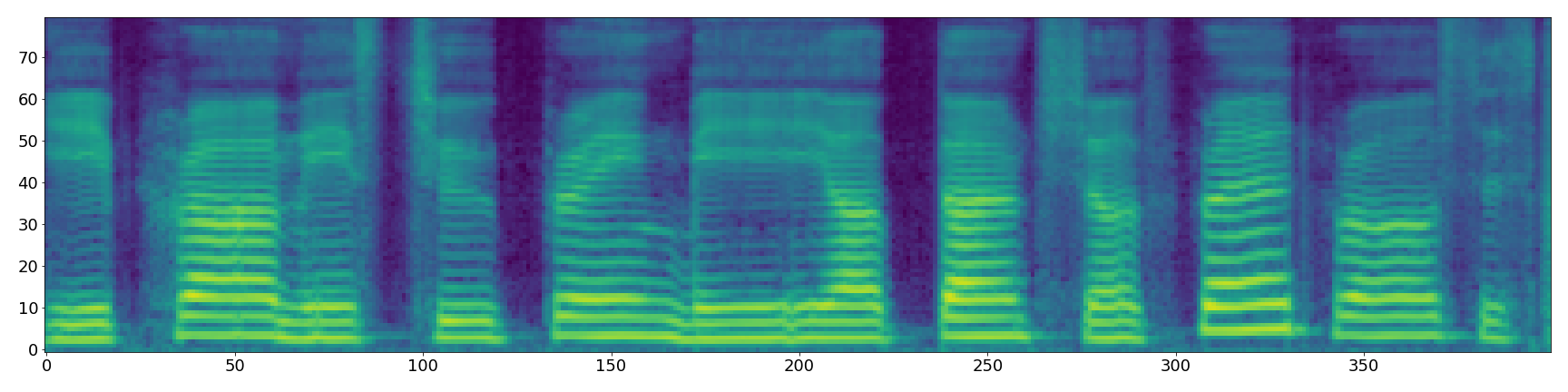}{c}{white}
         \phantomsubcaption\ignorespaces\label{speaker_c}
    \end{subfigure}%
    \vspace{-0.90\baselineskip}

    \vspace{-1.2\baselineskip}
    \caption{Mel spectrograms of the speaker conversion results.
    \textbf{(a)} is the input and the lyrics are "He promised to bring me a bunch of red roses." 
    \textbf{(b)-(c)} is the speaker conversion samples.
    The speaker embedding is switched and the pitches are multiplied by 1/2 to match the pitch range of the male target speakers.
    }
    \label{fig:control_speaker}
\end{figure*}

\clearpage

\section{Controlling with voice of the user.} \label{user_control}
\begin{figure*}[h!]
\setlength{\textfloatsep}{0pt}%

\newcommand{\incaptionimg}[3]{
  \begin{tikzpicture}[every node/.style={inner sep=0,outer sep=0}]
    \draw node[name=micrograph] {\includegraphics[width=\textwidth]{#1}}; 
    \draw  (micrograph.north west)  node[anchor=north west,yshift=-0.25cm,xshift=0.45cm,#3]{\textbf{\small{(#2)}}}; 
  \end{tikzpicture}
}

\captionsetup[subfigure]{labelformat=empty}
    \setlength{\textfloatsep}{0pt}%
    \setlength{\intextsep}{0pt}
    \centering
    \begin{subfigure}[b]{\linewidth}
    \subcaption{\textit{Author}}
    \end{subfigure}%
        \vspace{-0.90\baselineskip}
     \begin{subfigure}[b]{\linewidth}
         \incaptionimg{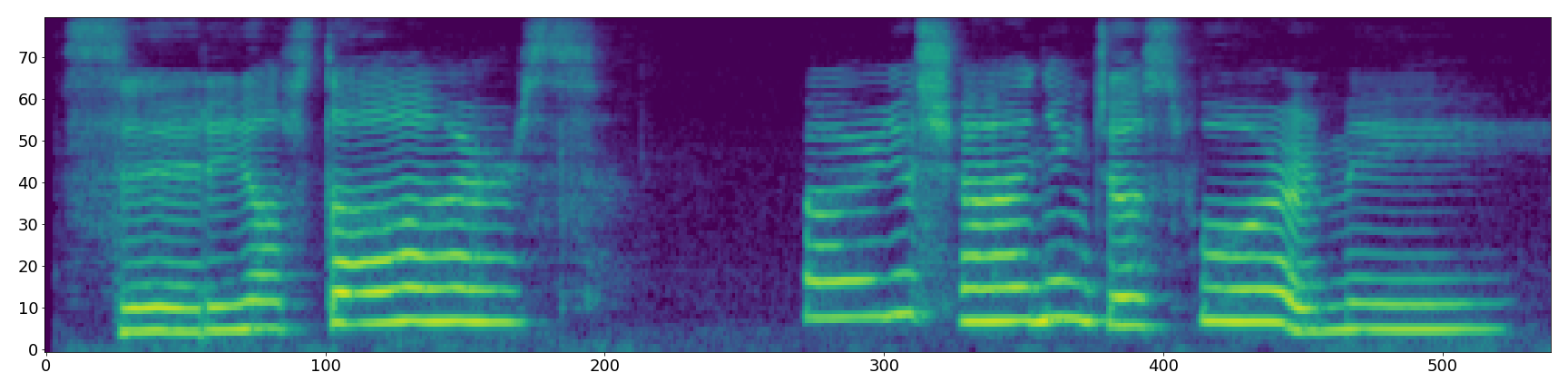}{a}{white}
         \phantomsubcaption\ignorespaces\label{user_a}
     \end{subfigure}%
    \vspace{-0.90\baselineskip}

    \begin{subfigure}[b]{\linewidth}
    \subcaption{\textit{MPOL}}
    \end{subfigure}%
    \vspace{-0.90\baselineskip}
    \begin{subfigure}[b]{\linewidth}
         \incaptionimg{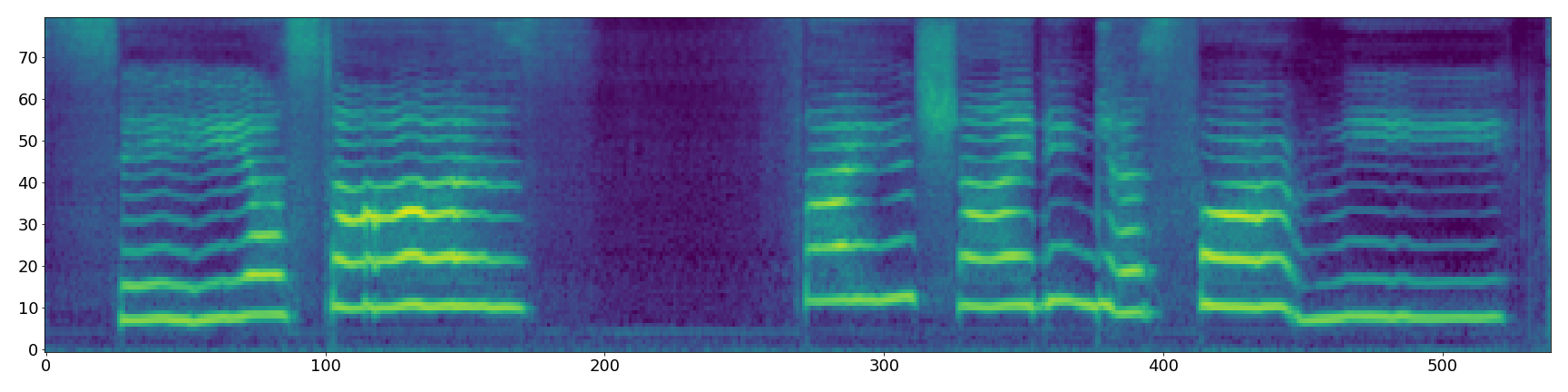}{b}{white}
         \phantomsubcaption\ignorespaces\label{user_b}
    \end{subfigure}%
    \vspace{-0.90\baselineskip}
    
    \begin{subfigure}[b]{\linewidth}
    \subcaption{\textit{Author} and \textit{MPOL} duet}
    \end{subfigure}%
    \vspace{-0.90\baselineskip}
    \begin{subfigure}[b]{\linewidth}
         \incaptionimg{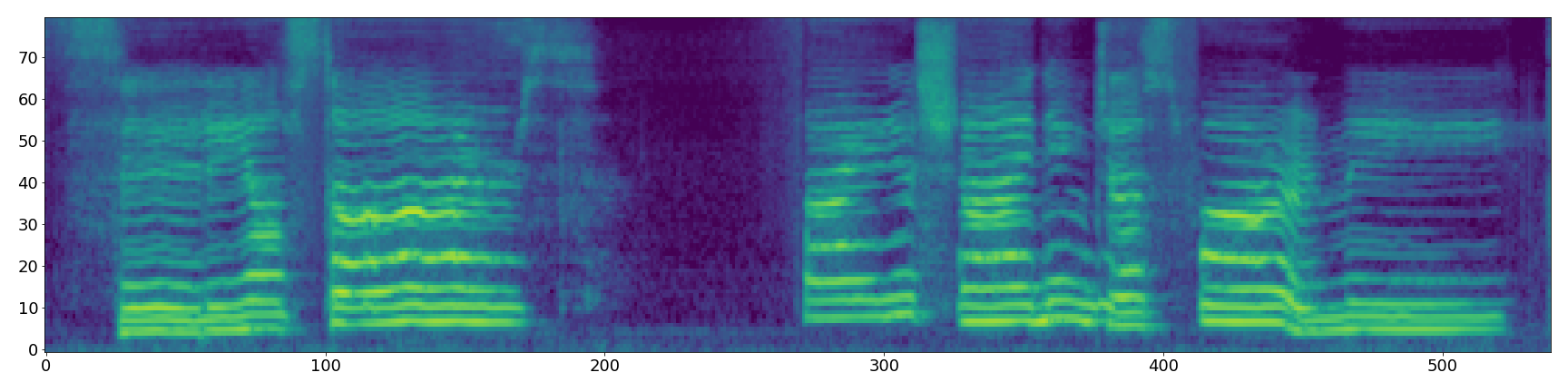}{c}{white}
         \phantomsubcaption\ignorespaces\label{user_c}
    \end{subfigure}%
    \vspace{-0.90\baselineskip}

    \vspace{-1.2\baselineskip}
    \caption{Mel spectrograms of the speaker conversion result and combined result with the original audio. The input audio is the author's voice and the lyrics are "City of stars are you shining just for me." The \mel of it is \textbf{(a)}. We switched the speaker embeddings and shifted the pitch to generate \textbf{(b)}. We combined \textbf{(a)} and \textbf{(b)} to generate a duet, \textbf{(c)}.
    }
    \label{fig:control_user}
\end{figure*}
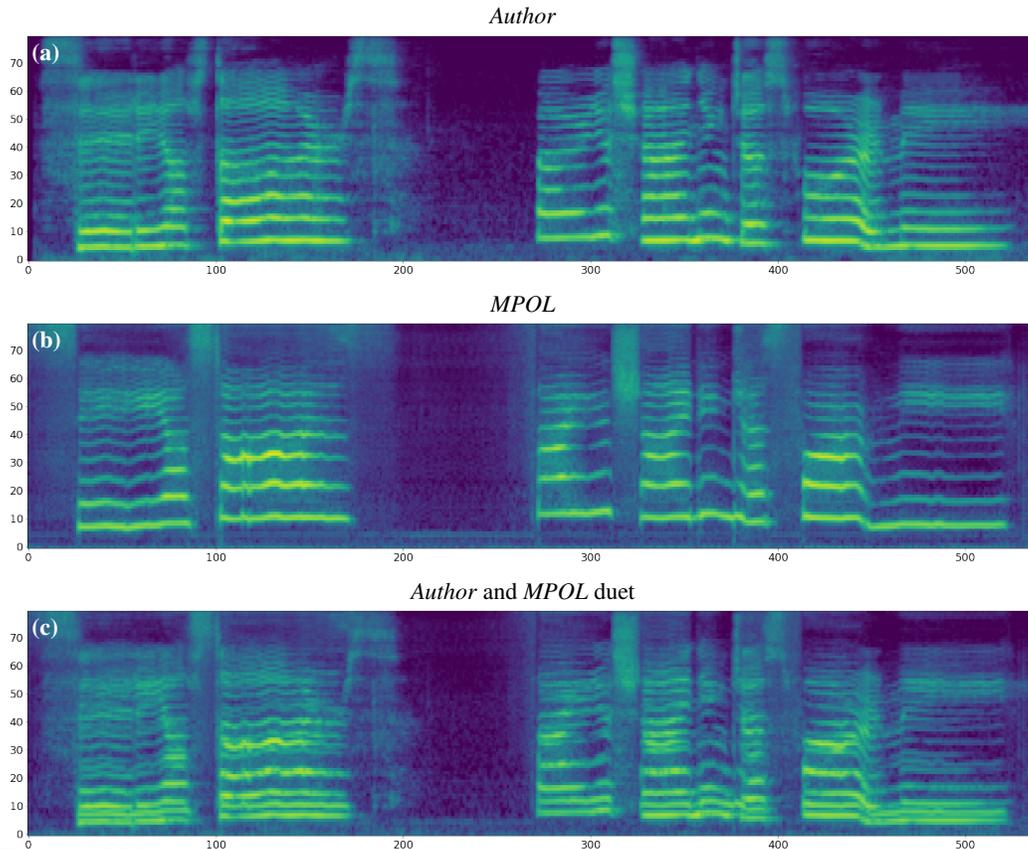

\section{Spectral Artifacts of HiFi-GAN} \label{artifacts}
\begin{figure*}[ht]
\setlength{\textfloatsep}{0pt}%

\newcommand{\incaptionimg}[3]{
  \begin{tikzpicture}[every node/.style={inner sep=0,outer sep=0}]
    \draw node[name=micrograph] {\includegraphics[width=\textwidth]{#1}}; 
    \draw  (micrograph.north west)  node[anchor=north west,yshift=-0.38cm,xshift=0.88cm,#3]{\textbf{\small{(#2)}}}; 
  \end{tikzpicture}
}

\captionsetup[subfigure]{labelformat=empty}
    \setlength{\textfloatsep}{0pt}%
    \setlength{\intextsep}{0pt}
    \centering
    \begin{subfigure}[b]{\linewidth}
    \subcaption{\textbf{Reference}}
    \end{subfigure}%
        \vspace{-0.90\baselineskip}
     \begin{subfigure}[b]{\linewidth}
         \incaptionimg{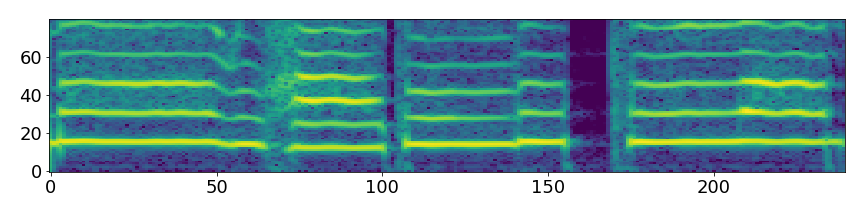}{a}{white}
         \phantomsubcaption\ignorespaces\label{artifacts_ref}
     \end{subfigure}%
    \vspace{-0.90\baselineskip}

    \begin{subfigure}[b]{\linewidth}
    \subcaption{\textbf{Our Model's Reconstruction}}
    \end{subfigure}%
    \vspace{-0.90\baselineskip}
    \begin{subfigure}[b]{\linewidth}
         \incaptionimg{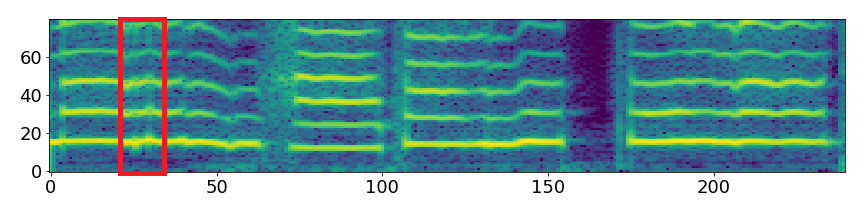}{b}{white}
         \phantomsubcaption\ignorespaces\label{artifacts_voc}
    \end{subfigure}%
    \vspace{-0.90\baselineskip}
    
    \begin{subfigure}[b]{\linewidth}
    \subcaption{\textbf{HiFi-GAN Reconstruction from the Ground-Truth Mel Spectrogram}}
    \end{subfigure}%
    \vspace{-0.90\baselineskip}
    \begin{subfigure}[b]{\linewidth}
         \incaptionimg{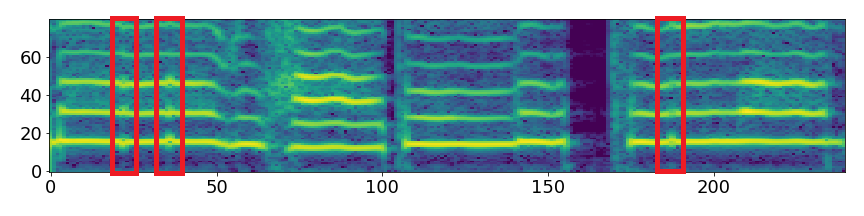}{c}{white}
         \phantomsubcaption\ignorespaces\label{artifacts_mod}
    \end{subfigure}%
    \vspace{-0.90\baselineskip}

    \vspace{-1.2\baselineskip}
    \caption{Spectral artifacts in low-frequency channels of linear spectrograms.
    \textbf{(a)} is the linear spectrogram of the input singing voice.
    \textbf{(b)} is the result of reconstructing the reference audio through our model. \textbf{(c)} is the reconstruction of HiFi-GAN from the ground-truth \mel.
    The artifacts are highlighted in red rectangles.}
    \label{fig:artifacts}
\end{figure*}
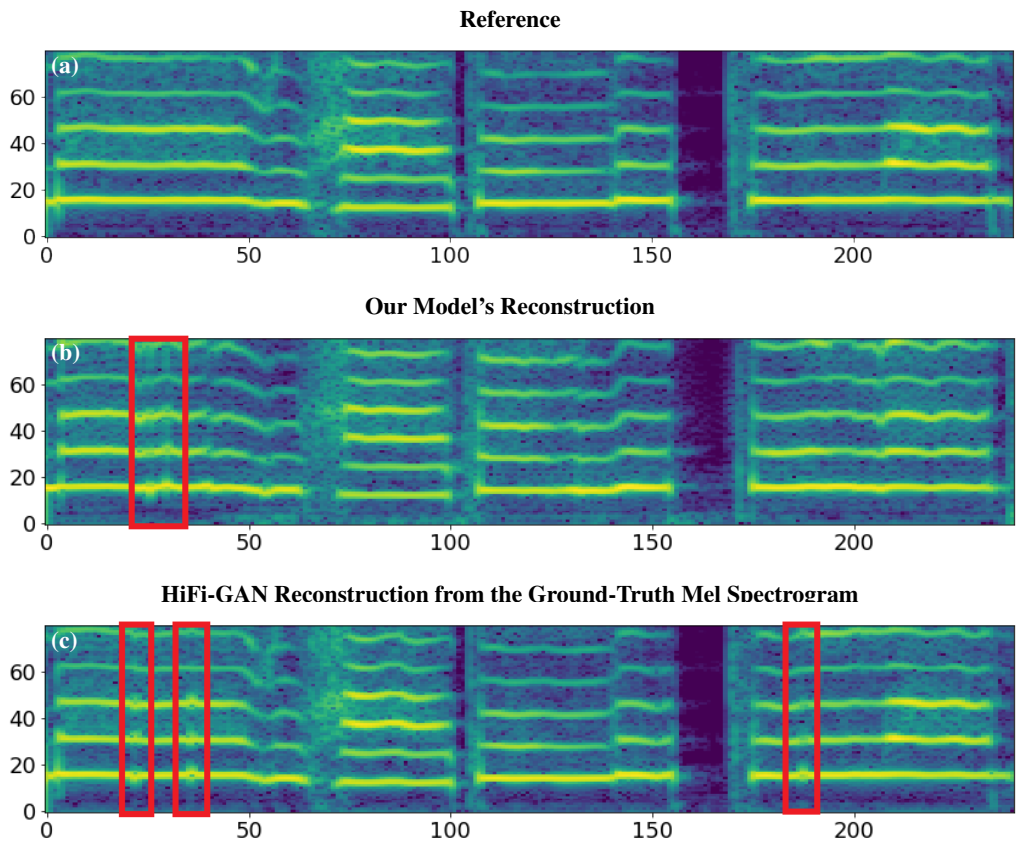

We observe that there are audible artifacts in our model's synthesized result, and it is also visible in spectrogram.
These noisy artifacts degrade the quality of the synthesized result of the model.
We also found that the similar audible artifacts is generated when the singing voice was reconstructed by HiFi-GAN.
Figure \ref{fig:artifacts} shows the comparison between the reference voice and the reconstructed voices with spectral artifacts.
We will resolve this issue in future works.

\end{document}